\documentclass{amsart}
\usepackage{amsfonts}

\usepackage{graphicx}
\usepackage[usenames]{color}

\input{tcilatex}
\input tcilatex

\begin{document}
\title[binomial catastrophes]{A generating function approach to Markov chains undergoing binomial
catastrophes}
\author{Branda Goncalves}
\author{Thierry Huillet}
\address{B. Goncalves and T. Huillet: Laboratoire de Physique Th\'{e}orique et
Mod\'{e}lisation, CY Cergy Paris Universit\'{e}, CNRS UMR-8089, 2 avenue
Adolphe-Chauvin, 95302 Cergy-Pontoise, FRANCE\\
E-mails: branda.goncalves@outlook.fr, Thierry.Huillet@cyu.fr }
\maketitle

\begin{abstract}
In a Markov chain population model subject to catastrophes, random
immigration events (birth), promoting growth, are in balance with the effect
of binomial catastrophes that cause recurrent mass removal (death). Using a
generating function approach, we study two versions of such population
models when the binomial catastrophic events are of a slightly different
random nature. In both cases, we describe the subtle balance between the two
birth and death conflicting effects.\newline



\end{abstract}

\section{Introduction, motivations and background}

In some simple growth process, some organism changes the amount of its
constitutive cells at random as follows. In the course of its lifetime, the
organism alternates at random busy and idle periods. In a busy growth
period, it produces a random or fixed amount of new cells which are being
added to its current stock of cells (say, the incremental or batch cells).
In an idle catastrophic period, the organism stands at risk being subject
say to virus/bacteria attacks or radiation, resulting in each of its
constitutive cells being either lysed or lost after mutations with some
fixed mortality probability, independently of each other. Under such a
catastrophic event, the population size is thus reduced according to a
binomial distribution with survival probability say $c$; hence the name
binomial catastrophes. It will happen that the current size of the organism
is reduced to zero at some random time. In a worst disaster scenario for
instance, all cells can be lysed in a single idle period leading
instantaneously to a first disastrous extinction event. From a first
extinction event, the organism can then either recover taking advantage of a
subsequent busy epoch and starting afresh from zero, or not, being stuck to
zero for ever. In the first case, we shall speak of a local extinction,
state zero being reflecting, while in the latter case eventual or global
extinction is at stake, state zero being absorbing.

As just described in words, the process under concern turns out to be a
Markov chain on the non-negative integers, displaying a subtle balance
between generalized birth and death events. Whenever it is ergodic, there
are infinitely many local extinctions and the pieces of sample paths
separating consecutive passages to zero are called excursions, the height
and length of which are of some relevance in the understanding of the
population size evolution. Excursions indeed form independent and
identically distributed (iid) blocks of this Markov chain.\newline

Stochastic models subject to catastrophes have a wide application in
different fields viz. bioscience, marketing, ecology, computer and natural
sciences, etc. For instance,

- In recent years, the study of catastrophe models has attracted much
attention due to their wide application in computer-communication networks
and digital telecommunication systems where interruptions due to various
types of virus attacks are referred to as catastrophes. Here, unfriendly
events, like virus attacks, result in abrupt changes in the state of the
system caused by the removal of some of its elements (packets), posing a
major threat to these queueing systems.\textbf{\ }Continuous-time queueing
models subject to total disasters have also been analyzed by a few
researchers such as \cite{BaS}, \cite{BE}\textbf{. }A detailed survey of
catastrophic events occurring in communication networks has been carried out
in \cite{Dab}.

- In oversimplistic earthquakes' physics models, the state of the process
may be viewed as the total energy embodied in the earth's crust system,
obtained as the sum of the energies of its constitutive blocks, each
carrying say a fixed unit quantum of energy. A busy period corresponds to an
accumulation of energy epoch in the system, while an idle period yields a
stress release event. More realistic growth/collapse or decay/surge models
in the same spirit, although in the continuum, were considered in \cite
{EK2006}, \cite{EK2007} and \cite{EK2009}, with physical applications in
mind, including stress release mechanisms in the physics of earthquakes.

- In forestry, a good management of the biomass depends on the prediction of
how steady growth periods alternate with tougher periods, due to the
occurrence of cataclysms such as hurricanes or droughts or floods, hitting
each tree in a similar way.

- In Economy, pomp periods often alternate with periods of scarcity when a
crisis equally strikes all the economic agents (the Joseph and Noah effects).%
\newline

Some aspects of related catastrophe models were recently addressed in \cite
{Bro}, \cite{Art}, \cite{Ec}, \cite{EcF} and \cite{CP}, chiefly in
continuous times. For instance, \cite{Ec}\textbf{\ }analyzed a
continuous-time binomial catastrophe model wherein individuals arrive
according to a compound Poisson process while catastrophes occur according
to a renewal process. He obtained the distribution of the population size at
post-catastrophe, arbitrary and pre-catastrophe epochs. In \cite{Art}, a
population model with the immigration process subject to binomial and
geometric catastrophes has been investigated; the authors obtained the
removed population size and maximum population size between two extinctions
(height of excursions). They also discussed the first extinction time and
the survival time of a tagged individual. In \cite{EcF} also, several
approaches for the transient analysis of a total disaster model were
introduced with an extension of the methodologies to the binomial
catastrophe model. The authors of \cite{Kap} carried out the transient
analysis of a binomial catastrophe model with birth/immigration-death
processes and obtained explicit expressions for factorial moments, which are
further used to develop the population size distribution. Results for the
equilibrium moments and population size distributions are also given.
Finally, recently, \cite{YP} obtained the stationary queue length of the $%
M^{X}/M/\mathbf{\infty }$ queue with binomial catastrophes in the heavy
traffic system, using a central limit theorem.

Let us finally mention that not all aspects of binomial catastrophes are
included in the above formulation. In \cite{Wil}, a density-dependent
catastrophe model of threatened population is investigated.
Density-dependence is shown here to have a significant effect on population
persistence (mean time to extinction), with a decreasing mean persistence
time at large initial population sizes and causing a relative increase at
intermediate sizes. A density-dependent model involving total disasters was
designed in \cite{Hui}.\newline

Discrete-time random population dynamics with catastrophes balanced by
random growth has a long history in the literature, starting with \cite
{Neuts}.\ Mathematically, binomial catastrophe models are Markov chains
(MCs) which are random walks on the non-negative integers (as a semigroup),
so differing from standard random walks on the integers (as a group) in that
a one-step move down from some positive integer cannot take one to a
negative state, resulting in transition probabilities being state-dependent, 
\cite{Dette}. Such MCs may thus be viewed as generalized birth and death
chains. The transient and equilibrium behaviour of such stochastic
population processes with either disastrous or mild binomial catastrophes is
one of the purposes of this work. We aim at studying the equilibrium
distribution of this process and deriving procedures for its approximate
computation. Another issue of importance concerns the measures of the risk
of extinction, first extinction time, time elapsed between two consecutive
extinction times and maximum population size reached in between.\newline

The detailed structure of the manuscript, attempting to realize this
program, can be summarized as follows:

$\bullet $ Section $2$ is designed to introduce the model in probabilistic
terms. We develop three particular important cases:

\begin{quote}
- Survival probability $c=1$. In this case, on a catastrophic event, the
chain remains in its current state with no depletion of cells at all.

- Survival probability $c=0$. This is a case of total disasters for which,
on a catastrophic event, the chain is instantaneously propelled to state $0$.

- the semi-stochastic growth/collapse scenario when the adjunction of
incremental cells, on a growth event, is deterministic being reduced to a
single element, and $c\neq \left\{ 0,1\right\} $ so that binomial mortality
is not degenerate.
\end{quote}

$\bullet $ In Section $3$, we discuss the conditions under which the chain
is recurrent (positive or null) or transient. We emphasize that positive
recurrence is generic, unless the batch random variables have unrealistic
very heavy tails. We use a generating function approach which is well-suited
to this analysis. When, as in the positive recurrent case, it is non
trivial, we discuss the shape of the invariant probability mass function of
the chain. Specifically,

\begin{quote}
- When $c\in \left[ 0,1\right) $, we show that the invariant probability
mass function is the one of a compound Poisson (infinitely divisible) random
variable, which is moreover discrete self-decomposable. As a result, it is
unimodal and we give conditions under which the mode is located at $0.$ We
refer to \cite{Steu} for the notions of discrete infinite divisibility
(compounding Poisson), self-decomposability and unimodality.

- When $c=0$, (total disasters), the invariant probability mass function is
the one of a geometric sum of the incremental (batch) random variables, so
in the compound Poisson class, but not necessarily self-decomposable. We
give some sufficient conditions under which it is self-decomposable, so
unimodal and then with mode necessarily at the origin.
\end{quote}

$\bullet $ In Section $4$, we analyze some important characteristics of the
chain in the positive recurrent regime, making use of its Green (potential)
kernel, including:

\begin{quote}
- from the Green (potential) kernel at $\left( 0,0\right) $: the contact
probability at state $0$ and first return time to $0$ (length of an
excursion).

- from the Green (potential) kernel at $\left( x_{0},0\right) $: the first
extinction time at $0$ when starting from $x_{0}>0.$ We show that this
random variable has geometric tails. Some representation results on the mean
(persistence) time to extinction are supplied.

- from the Green (potential) kernel at $\left( x_{0},y_{0}\right) $: first
hitting time of state $y_{0}$ when starting from $x_{0}.$

- Average position of the chain started at $x_{0}>0$ (transient analysis).

- Height of an excursion making use of the idea of the scale or harmonic
sequence of the chain.
\end{quote}

$\bullet $ In Section $5$, we sketch some brief insight on the way the
parameters of the chain could be estimated from an $N-$sample of
observations.

$\bullet $ In Section $6$, we study an absorbed version of the binomial
catastrophe chain in view of further analyzing the quasi-distribution of the
process conditioned on being currently not extinct.

$\bullet $ Finally, a variant of the binomial catastrophe model is
introduced and studied in Section $7$. It was motivated by the following
observation: suppose at each step of its evolution, the size of some
organism either grows by a random number of batch cells (not reduced to a
fixed number) or shrinks deterministically by only \textbf{one} unit (a
semi-stochastic decay/surge scenario `dual' to the semi-stochastic
growth/collapse one). The above standard binomial catastrophe model is not
able to represent a scenario where at least one individual is removed from
the population at catastrophic events. To remedy this, we therefore define
and analyze a variant of the standard binomial model where, on a
catastrophic event, the binomial shrinking mechanism applies to all
currently alive cells but one which is systematically removed. We also take
control on the future of the population once it hits state zero. Despite the
apparent small changes in the definition of this modified binomial
catastrophe Markov chain, the impact on its asymptotic behavior is shown to
be drastic.

\section{The binomial catastrophe model}

We first describe the model as a toy model one.

\subsection{The model}

Consider a discrete-time MC $X_{n}$ taking values in $\Bbb{N}_{0}:=\left\{
0,1,2,...\right\} .$ With $b_{n}\left( c\right) $, $n=1,2,...$ an
independent identically distributed (iid) sequence of Bernoulli random
variables (rvs) with success parameter $c\in \left( 0,1\right) $, let $%
c\circ X_{n}=\sum_{m=1}^{X_{n}}b_{m}\left( c\right) $ denote the Bernoulli
thinning of $X_{n}$\footnote{%
In words, the thinning operation acting on a discrete random variable is the
natural discrete analog of scaling a continuous variable, i.e., multiplying
it by a constant in $\left[ 0,1\right] $.}. Let $\beta _{n}$, $n=1,2,...$ be
an iid birth sequence of rvs with values in $\Bbb{N}=\left\{ 1,2,...\right\} 
$. The dynamics of the MC under concern here is a balance between birth and
death events according as ($p+q=1$): 
\begin{eqnarray*}
X_{n+1} &=&1\circ X_{n}+\beta _{n+1}=X_{n}+\beta _{n+1}\text{ with
probability }p \\
X_{n+1} &=&c\circ X_{n}\text{ with probability }q=1-p.
\end{eqnarray*}
This model was considered by \cite{Neuts}, \cite{BRS}. We have $c\circ
X_{n}\sim $bin$\left( X_{n},c\right) $ hence the name binomial catastrophe.
The binomial effect is appropriate when,\textbf{\ }on a catastrophic event,
the individuals of the current population each die (with probability $1-c$)
or survive (with probability $c$)\textbf{\ }in an independent and even way,
resulting in a drastic depletion of individuals at each step. Owing to: $%
c\circ X_{n}=X_{n}-$ $\left( 1-c\right) \circ X_{n}$, the number of stepwise
removed individuals is $\left( 1-c\right) \circ X_{n}$ with probability (wp) 
$q.$ This way of depleting the population size (at shrinkage times) by a
fixed random fraction $c$ of its current size is very drastic, especially if 
$X_{n}$ happens to be large. Unless $c$ is very close to $1$ in which case
depletion is modest (the case $c=1$ is discussed below), it is very unlikely
that the size of the upward moves will be large enough to compensate
depletion while producing a transient chain drifting at $\infty $. We will
make this very precise below.

Note also $X_{n}=0\Rightarrow X_{n+1}=\beta _{n+1}$ wp $p$, $X_{n+1}=0$ wp $%
q $ (translating reflection of $X_{n}$ at $0$ if $q>0$)$.$\newline

Let $B_{n}\left( p\right) $, $n=1,2,...$ be an iid sequence of Bernoulli rvs
with $\mathbf{P}\left( B_{1}\left( p\right) =1\right) =p$, $\mathbf{P}\left(
B_{1}\left( p\right) =0\right) =q.$ Let $C_{n}\left( p,c\right) =B_{n}\left(
p\right) +c\overline{B}_{n}\left( p\right) =c^{\overline{B}_{n}\left(
p\right) }$ where $\overline{B}_{n}\left( p\right) =1-B_{n}\left( p\right) $%
. The above process' dynamics (driven by $B_{n}\left( p\right) \beta _{n}$)
is compactly equivalent to 
\begin{eqnarray*}
X_{n+1} &=&c^{\overline{B}_{n+1}\left( p\right) }\circ X_{n}+B_{n+1}\left(
p\right) \beta _{n+1},\text{ }X_{0}=0 \\
&=&\overline{B}_{n+1}\left( p\right) \cdot c\circ X_{n}+B_{n+1}\left(
p\right) \left( X_{n}+\beta _{n+1}\right) ,
\end{eqnarray*}
$\left( X_{n},\beta _{n+1}\right) $ being mutually independent. The thinning
coefficients are now $c^{\overline{B}_{n+1}\left( p\right) },$ so random.%
\newline

With $b_{x}:=\mathbf{P}\left( \beta =x\right) $, $x\geq 1$ and $d_{x,y}:=%
\binom{x}{y}c^{y}\left( 1-c\right) ^{x-y}$ the binomial probability mass
function (pmf), the one-step-transition matrix $P$ of the MC $X_{n}$ is
given by: 
\begin{equation*}
P\left( 0,0\right) =q,P\left( 0,y\right) =p\mathbf{P}\left( \beta =y\right)
=pb_{y},y\geq 1
\end{equation*}
\begin{equation}
P\left( x,y\right) =q\binom{x}{y}c^{y}\left( 1-c\right) ^{x-y}=qd_{x,y}\text{%
, }x\geq 1\text{ and }0\leq y\leq x  \label{P}
\end{equation}
\begin{equation*}
P\left( x,y\right) =p\mathbf{P}\left( \beta =y-x\right) =pb_{y-x}\text{, }%
x\geq 1\text{ and }y>x.
\end{equation*}
If $\beta $ has first and second moment finite, with $\overline{c}=1-c$, $%
\overline{c}\circ x\sim $bin$\left( x,\overline{c}\right) $, as $x$ gets
large 
\begin{eqnarray*}
m_{1}\left( x\right) &=&\mathbf{E}\left( \left( X_{n+1}-X_{n}\right) \mid
X_{n}=x\right) =p\mathbf{E}\left( \beta \right) -q\overline{c}x\sim -q%
\overline{c}x \\
m_{2}\left( x\right) &=&\mathbf{E}\left( \left( X_{n+1}-X_{n}\right)
^{2}\mid X_{n}=x\right) =p\mathbf{E}\left( \beta ^{2}\right) +q\left( 
\overline{c}x+\overline{c}^{2}x\left( x-1\right) \right) \sim q\overline{c}%
^{2}x^{2}
\end{eqnarray*}
with 
\begin{equation*}
\frac{m_{1}\left( x\right) }{m_{2}\left( x\right) }\sim -\frac{1}{\overline{c%
}x},\text{ }x\text{ large.}
\end{equation*}
Note the variance of the increment is 
\begin{equation*}
\sigma ^{2}\left( X_{n+1}-X_{n}\mid X_{n}=x\right) =p\sigma ^{2}\left( \beta
\right) +qc\overline{c}x\sim qc\overline{c}x.
\end{equation*}

\subsection{Special cases}

$\left( i\right) $ When $c=1$, the lower triangular part of $P$ vanishes
leading to 
\begin{eqnarray*}
P\left( 0,0\right) &=&q\text{, }P\left( 0,y\right) =p\mathbf{P}\left( \beta
=y\right) =pb_{y}\text{, }y\geq 1 \\
P\left( x,y\right) &=&0\text{, }x\geq 1\text{ and }0\leq y<x\text{, }P\left(
x,x\right) =q\text{, }x\geq 1 \\
P\left( x,y\right) &=&p\mathbf{P}\left( \beta =y-x\right) =pb_{y-x}\text{, }%
x\geq 1\text{ and }y>x.
\end{eqnarray*}
The transition matrix $P$ is upper-triangular with diagonal terms. The
process $X_{n}$ is non-decreasing, so it drifts to $\infty $.\newline

$\left( ii\right) $ When $c=0$ (total disasters), 
\begin{eqnarray*}
P\left( 0,0\right) &=&q\text{, }P\left( 0,y\right) =p\mathbf{P}\left( \beta
=y\right) =pb_{y}\text{, }y\geq 1 \\
P\left( x,y\right) &=&0\text{, }x\geq 1\text{ and }0<y\leq x\text{, }P\left(
x,0\right) =q\text{, }x\geq 1 \\
P\left( x,y\right) &=&p\mathbf{P}\left( \beta =y-x\right) =pb_{y-x}\text{, }%
x\geq 1\text{ and }y>x.
\end{eqnarray*}
When a downward move occurs, it takes instantaneously $X_{n}$ to zero (a
case of total disasters), independently of the value of $X_{n}$. This means
that, defining $\tau _{x_{0},0}=\inf \left( n\geq 1:X_{n}=0\mid
X_{0}=x_{0}\right) $, the first extinction time of $X_{n}$, $\mathbf{P}%
\left( \tau _{x_{0},0}=x\right) =qp^{x-1}$, $x\geq 1$, a geometric
distribution with success parameter $q$, with mean $\mathbf{E}\left( \tau
_{x_{0},0}\right) =1/q$, independently of $x_{0}\geq 1.$ Note that $\tau
_{0,0}$, as the length of any excursion between consecutive visits to $0,$
also has a geometric distribution with success parameter $q$ and finite mean 
$1/q$. In addition, the height $H$ of an excursion is clearly distributed
like $\sum_{x=1}^{\tau _{0,0}-1}\beta _{x}$ (with the convention $%
\sum_{x=1}^{0}\beta _{x}=0$)$.$ Finally, this particular Markov chain
clearly is always positive recurrent, whatever the distribution of $\beta $.
Consecutive excursions are the iid pieces of this random walk on the
non-negative integers.

Some Markov catastrophe models involving total disasters are described in 
\cite{Sw}, \cite{Hui}. \newline

$\left( iii\right) $ If $\beta \sim \delta _{1}$ a move up results in the
addition of only one individual, which is the simplest deterministic drift
upwards. In this case, the transition matrix $P$ is lower- Hessenberg. This
model constitutes a simple discrete version of a semi-stochastic
growth/collapse model in the continuum, \cite{EK2006}.\newline

\section{Recurrence versus transience}

Using a generating function approach, we start with the transient analysis
before switching to the question of equilibrium.

\subsection{The transient analysis}

Let $B\left( z\right) =\mathbf{E}\left( z^{\beta }\right) $ be the pgf of $%
\beta =\beta _{1}$, as an absolutely monotone function on $\left[ 0,1\right] 
$\footnote{%
A function $B$ is said to be absolutely monotone on $\left( 0,1\right) $ if
it has all its derivatives $B^{\left( n\right) }\left( z\right) \geq 0$ for
all $z\in \left( 0,1\right) $. Pgfs are absolutely monotone and the
composition of two pgfs is a pgf.}. Let $\mathbf{\pi }_{n}^{\prime }=\left(
\pi _{n}\left( 0\right) ,\pi _{n}\left( 1\right) ,...\right) $ where $\pi
_{n}\left( x\right) =\mathbf{P}_{0}\left( X_{n}=x\right) $ and $^{\prime }$
denotes the transposition. With $\mathbf{z}=\left( 1,z,z^{2},...\right)
^{\prime }$, a column vector obtained as the transpose $^{\prime }$ of the
row vector $\left( 1,z,z^{2},...\right) $, define

\begin{equation*}
\Phi _{n}\left( z\right) =\mathbf{E}_{0}\left( z^{X_{n}}\right) =\mathbf{\pi 
}_{n}^{\prime }\mathbf{z,}
\end{equation*}
the pgf of $X_{n}$. With $D_{\mathbf{z}}=$diag$\left( 1,z,z^{2},...\right) $%
, $\mathbf{z}=D_{\mathbf{z}}\mathbf{1}$, the time evolution $\mathbf{\pi }%
_{n+1}^{\prime }=\mathbf{\pi }_{n}^{\prime }P$ yields 
\begin{equation*}
\Phi _{n+1}\left( z\right) =\mathbf{\pi }_{n+1}^{\prime }\mathbf{z=\pi }%
_{n}^{\prime }P\mathbf{z}=\mathbf{\pi }_{n}^{\prime }PD_{\mathbf{z}}\mathbf{%
1,}
\end{equation*}
leading to the transient dynamics 
\begin{equation}
\Phi _{n+1}\left( z\right) =pB\left( z\right) \Phi _{n}\left( z\right)
+q\Phi _{n}\left( 1-c\left( 1-z\right) \right) \text{, }\Phi _{0}\left(
z\right) =1.  \label{R}
\end{equation}
The fixed point pgf of $X_{\infty },$ if it exists, solves 
\begin{equation}
\Phi _{\infty }\left( z\right) =pB\left( z\right) \Phi _{\infty }\left(
z\right) +q\Phi _{\infty }\left( 1-c\left( 1-z\right) \right) \text{.}
\label{Fix}
\end{equation}
\newline

$\left( i\right) $ When $c=1$, there is no move down possible. The only
solution to $\Phi _{\infty }\left( z\right) =pB\left( z\right) \Phi _{\infty
}\left( z\right) +q\Phi _{\infty }\left( z\right) $ is $\Phi _{\infty
}\left( z\right) =0,$ corresponding to $X_{\infty }\sim \delta _{\infty }.$
Indeed, when $c=1$, combined to $\Phi _{\infty }\left( 1\right) =1,$%
\begin{eqnarray*}
\Phi _{n+1}\left( z\right) &=&\left( q+pB\left( z\right) \right) \Phi
_{n}\left( z\right) ,\Phi _{0}\left( z\right) =1 \\
\Phi _{n}\left( z\right) ^{1/n} &=&q+pB\left( z\right) ,
\end{eqnarray*}
showing that, if $1\leq \rho :=B^{\prime }\left( 1\right) =\mathbf{E}\left(
\beta \right) <\infty $, $n^{-1}X_{n}\rightarrow q+p\rho \geq 1$ almost
surely as $n\rightarrow \infty $. The process $X_{n}$ is transient in that,
after a finite number of passages in state $0$, it drifts to $\infty $.%
\newline

$\left( ii\right) $ When $c=0$ (total disasters), combined to $\Phi _{\infty
}\left( 1\right) =1$, (\ref{Fix}) yields 
\begin{equation}
\Phi _{\infty }\left( z\right) =\frac{q}{1-pB\left( z\right) }=:\phi
_{\Delta }\left( z\right) ,  \label{Del}
\end{equation}
as an admissible pgf solution. In (\ref{Del}), we introduced the
integral-valued rv $\Delta $ whose pgf is $\phi _{\Delta }\left( z\right) =%
\mathbf{E}\left( z^{\Delta }\right) =q/\left( 1-pB\left( z\right) \right) $
obtained while compounding a shifted-geometric pgf $q/\left( 1-pz\right) $
with the pgf $B\left( z\right) $ of the $\beta $s \footnote{%
A geometric$\left( q\right) $ rv with success probability $q$ takes values
in $\Bbb{N=}\left\{ 1,2,...\right\} $. A shifted geometric$\left( q\right) $
rv with success probability $q$ takes values in $\Bbb{N}_{0}=\left\{
0,1,2,...\right\} $. It is obtained while shifting the former one by one
unit.}$.$ We conclude that in the total disaster setup when $c=0$, the law
of $X_{\infty }$ is obtained as a compound shifted-geometric sum $\Delta $
of the $\beta $s, whatever the distribution of $\beta $. Note that the rv $%
\Delta $ clearly is the height of any total disaster excursion, as the
sample path between any two consecutive visits of $X_{n}$ to $0$. The length
of such excursions clearly is geometric with success probability $q$, in
that case.

\subsection{Existence and shape of an invariant pmf ($c\in \left[ 0,1\right) 
$)}

We shall distinguish two cases.

$\bullet $ \textbf{The case }$c\in \left( 0,1\right) .$

From (\ref{Fix}) and (\ref{Del}), the limit law pgf $\Phi _{\infty }\left(
z\right) $, if it exists, solves the functional equation 
\begin{equation}
\Phi _{\infty }\left( z\right) =\phi _{\Delta }\left( z\right) \Phi _{\infty
}\left( 1-c\left( 1-z\right) \right) ,  \label{FE}
\end{equation}
so that, formally 
\begin{equation}
\Phi _{\infty }\left( z\right) =\prod_{n\geq 0}\phi _{\Delta }\left(
1-c^{n}\left( 1-z\right) \right) ,  \label{prod}
\end{equation}
as an infinite product pgf.\newline

\textbf{Proposition}. \emph{The invariant measure exists for all }$c\in
\left( 0,1\right) $\emph{\ if and only if }$\mathbf{E}\left( \log _{+}\Delta
\right) <\infty .$

\emph{Proof} (Theorem $2$ in \cite{Neuts}): By a comparison argument, we
need to check the conditions under which $\pi \left( 0\right) =\Phi _{\infty
}\left( 0\right) $ converges to a positive number. We get 
\begin{eqnarray*}
\Phi _{\infty }\left( 0\right) &=&\prod_{n\geq 0}\phi _{\Delta }\left(
1-c^{n}\right) >0\Leftrightarrow \sum_{n\geq 0}\left( 1-\phi _{\Delta
}\left( 1-c^{n}\right) \right) <\infty \\
&\Leftrightarrow &\int_{0}^{1}\frac{1-\phi _{\Delta }\left( z\right) }{1-z}%
dz<\infty \Leftrightarrow \sum_{x\geq 1}\log x\mathbf{P}\left( \Delta
=x\right) =\mathbf{E}\left( \log _{+}\Delta \right) <\infty ,
\end{eqnarray*}
meaning that $\Delta $ has a finite logarithmic first moment.

For most $\beta $s therefore, the process $X_{n}$ is positive recurrent, in
particular if $\beta $ has finite mean.\newline

When $\beta $ has finite first and second order moments, so do $\Delta $ and 
$X_{\infty }$ which exist. Indeed:

If $B^{\prime }\left( 1\right) =\mathbf{E}\beta =\rho <\infty ,$ (with $%
\mathbf{E}\Delta =\left( p\rho \right) /q$) 
\begin{equation*}
\Phi _{\infty }^{\prime }\left( 1\right) =q\left( \frac{p\rho }{q^{2}}+\frac{%
c}{q}\Phi _{\infty }^{\prime }\left( 1\right) \right) \Rightarrow \Phi
_{\infty }^{\prime }\left( 1\right) =\mathbf{E}\left( X_{\infty }\right)
=:\mu =\frac{p\rho }{q\left( 1-c\right) }<\infty
\end{equation*}

If $B^{^{\prime \prime }}\left( 1\right) <\infty $, $X_{\infty }$ has finite
variance: 
\begin{equation*}
\Phi _{\infty }^{^{\prime \prime }}\left( 1\right) -\Phi _{\infty }^{\prime
}\left( 1\right) ^{2}=\frac{p}{q}\frac{1}{1-c^{2}}\left( B^{^{\prime \prime
}}\left( 1\right) +2\frac{p}{q}\frac{2c-1}{1-c}\rho ^{2}\right)
\end{equation*}
\newline

\textbf{Counter-example:} With $\beta ,C>0$, suppose that $\mathbf{P}\left(
\beta >x\right) \sim _{x\uparrow \infty }C\left( \log x\right) ^{-\beta }$\
translating that $\beta $\ has very heavy logarithmic tails (any other than
logarithmic slowly varying function would do the job as well). Then $\mathbf{%
E}\beta ^{q}=\infty $\ for all $q>0$\ and $\beta $\ has no moments of
arbitrary positive order. Equivalently, $B\left( z\right) \sim _{z\downarrow
1}1-\frac{C}{\left( -\log \left( 1-z\right) \right) ^{\beta }}$. Therefore,
with $C^{\prime }=pC/q,$%
\begin{equation*}
\phi _{\Delta }\left( z\right) =\frac{q}{1-pB\left( z\right) }\sim
_{z\downarrow 1}1-\frac{C^{\prime }}{\left( -\log \left( 1-z\right) \right)
^{\beta }}
\end{equation*}
translating that $\mathbf{P}\left( \Delta >y\right) \sim _{y\uparrow \infty
}C^{\prime }\left( \log y\right) ^{-\beta }$\ shares the same tail behavior
as $\beta $. From this, $\mathbf{P}\left( \log \Delta >x\right) \sim
_{j\uparrow \infty }C^{\prime }x^{-\beta }$\ so that $\log \Delta $\ has a
first moment if and only if $\beta >1.$\ For such a (logarithmic tail) model
of $\beta $, we conclude that $X$\ remains positive recurrent if $\beta >1$\
and starts being transient only if $\beta <1.$\ The case $\beta =1$\ is a
critical null-recurrent situation.

Being strongly attracted to $0$, the binomial catastrophe model exhibits a
recurrence/transience transition but only for such very heavy-tailed choices
of $\beta $. Recall that:

When positive recurrent, the chain visits state $0$ infinitely often and the
expected return time to $0$ has finite mean.

When null recurrent, the chain visits state $0$ infinitely often but the
expected return time to $0$ has infinite mean.

When transient, the chain visits state $0$ a finite number of times before
drifting to $\infty $ for ever after an infinite number of steps (no finite
time explosion is possible for discrete-time Markov chains).\newline

\textbf{Corollary.} \emph{If the process is null recurrent or transient, no
non-trivial (}$\neq \mathbf{0}^{\prime }$\emph{) invariant measure exists.}

\emph{Proof:} This is because\textbf{, }$\Phi _{\infty }\left( z\right) $
being an absolutely monotone function on $\left[ 0,1\right] $ if it exists, 
\begin{equation*}
\pi \left( 0\right) =\Phi _{\infty }\left( 0\right) =0\Rightarrow \Phi
_{\infty }\left( 1-c\right) =0\Rightarrow \pi \left( x\right) =0\text{ for
all }x\geq 1.
\end{equation*}
\newline

\textbf{Clustering} (sampling at times when thinning occurs, time change):
Let $G=\inf \left( n\geq 1:B_{n}\left( p\right) =0\right) $, with $\mathbf{P}%
\left( G=k\right) =p^{k-1}q$, $\mathbf{E}\left( z^{G-1}\right) =\frac{q}{1-pz%
}$. $G$ is the time elapsed between two consecutive catastrophic events. So
long as there is no thinning of $X$ (a catastrophic event), the process
grows of $\Delta =\sum_{k=1}^{G-1}\beta _{k}$ individuals. Consider a
time-changed process of $X$ whereby one time unit is the time elapsed
between consecutive catastrophic events. During this laps of time, the
original process $X_{n}$ grew of $\Delta $ individuals, before shrinking to
a random amount of its current size at catastrophe times. We are thus led to
consider the time-changed integral-valued Ornstein-Uhlenbeck process \textbf{%
[}also known as an Integer-Valued Autoregressive of Order $1$ (in short
INAR(1)) process\textbf{, }see\textbf{\ }\cite{McK}\textbf{]:}

\begin{equation*}
X_{k+1}=c\circ X_{k}+\Delta _{k+1},\text{ }X_{0}=0,
\end{equation*}
with $\Delta _{k}$, $k=1,2,...$ an iid sequence of compound shifted
geometric rvs$.$

In this form, $X_{k}$ is also a pure-death subcritical branching process
with immigration, $\Delta _{k+1}$ being the number of immigrants at
generation $k+1$, independent of $X_{k}.$ With $\Phi _{k}\left( z\right) =%
\mathbf{E}\left( z^{X_{k}}\right) $, we have 
\begin{equation*}
\Phi _{k+1}\left( z\right) =\frac{q}{1-pB\left( z\right) }\Phi _{k}\left(
1-c\left( 1-z\right) \right) \text{, }\Phi _{0}\left( z\right) =1.
\end{equation*}
The limit law (if it exists) $\Phi _{\infty }\left( z\right) $ also solves (%
\ref{FE}). Thus 
\begin{equation*}
\Phi _{\infty }\left( z\right) =\prod_{n\geq 0}\phi _{\Delta }\left(
1-c^{n}\left( 1-z\right) \right) ,
\end{equation*}
corresponding to 
\begin{equation*}
X_{\infty }\overset{d}{=}\sum_{n\geq 0}c^{n}\circ \Delta _{n+1}.
\end{equation*}
As conventional wisdom suggests, the time-changed process has the same limit
law as the original binomial catastrophe model, so if and only if the
condition $\mathbf{E}\log _{+}\Delta <\infty $ holds.\newline

\textbf{Proposition. }\emph{When the law of }$X_{\infty }$\emph{\ exists, it
is discrete self-decomposable (SD).}

\emph{Proof:} This follows, for example, from\textbf{\ }Theorem $3.1$ of%
\textbf{\ }\cite{BS} and the INAR(1) process representation of $\left(
X_{k}\right) $. See \cite{Steu} for an account on discrete SD distributions,
as a remarkable subclass of compound Poisson ones.

The rv $X_{\infty }$ being SD, it is unimodal, with mode at the origin if $%
\pi \left( 1\right) <\pi \left( 0\right) $, or with two modes at $\left\{
0,1\right\} $ if $\frac{\pi \left( 1\right) }{\pi \left( 0\right) }=1$ (see 
\cite{SH}, Theorem $4.20$).

With $\mathbf{P}\left( \Delta =1\right) =\phi _{\Delta }^{\prime }\left(
0\right) =pq\mathbf{P}\left( \beta =1\right) $, we have

\begin{eqnarray*}
\pi \left( 0\right) &=&\Phi _{\infty }\left( 0\right) =q\Phi _{\infty
}\left( 1-c\right) \\
\pi \left( 1\right) &=&\Phi _{\infty }^{\prime }\left( 0\right) =\mathbf{P}%
\left( \Delta =1\right) \Phi _{\infty }\left( 1-c\right) +qc\Phi _{\infty
}^{\prime }\left( 1-c\right) \\
&=&\frac{\mathbf{P}\left( \Delta =1\right) }{q}\pi \left( 0\right) +qc\Phi
_{\infty }^{\prime }\left( 1-c\right) >p\mathbf{P}\left( \beta =1\right) \pi
\left( 0\right)
\end{eqnarray*}
A condition for unimodality at $0$ is thus 
\begin{equation}
\left( \log \Phi _{\infty }\right) ^{\prime }\left( 1-c\right) <\frac{1-p%
\mathbf{P}\left( \beta =1\right) }{c}.  \label{uni}
\end{equation}
Note also 
\begin{eqnarray*}
\pi \left( 1\right) &=&\Phi _{\infty }^{\prime }\left( 0\right) =\sum_{m\geq
0}c^{m}\phi _{\Delta }^{\prime }\left( 1-c^{m}\right) \prod_{n\neq m}\phi
_{\Delta }\left( 1-c^{n}\right) \\
&=&\pi \left( 0\right) \sum_{m\geq 0}c^{m}\left( \log \phi _{\Delta }\right)
^{\prime }\left( 1-c^{m}\right) \\
&=&\pi \left( 0\right) \sum_{m\geq 0}c^{m}\frac{pB^{\prime }\left(
1-c^{m}\right) }{1-pB\left( 1-c^{m}\right) }
\end{eqnarray*}
giving a closed-form condition for unimodality at $0$. For instance, if $%
B\left( z\right) =z,$ $\pi \left( 1\right) <\pi \left( 0\right) $ if and
only if 
\begin{equation*}
\sum_{m\geq 0}\frac{pc^{m}}{q+pc^{m}}<1.
\end{equation*}

\textbf{Tails of }$X_{\infty }$\textbf{.} The probabilities $\pi \left(
x\right) =\left[ z^{x}\right] \Phi _{\infty }\left( z\right) $, $x\geq 1$
(the $z^{x}-$coefficient in the power series expansion of $\Phi _{\infty
}\left( z\right) $) are hard to evaluate. However, some information on the
large $x$ tails $\sum_{y>x}\pi \left( y\right) $ can be estimated in some
cases.

- Consider a case where $B\left( z\right) \sim C\cdot \left(
1-z/z_{c}\right) ^{-1}$ as $z\rightarrow z_{c}>1$ so that $\mathbf{P}\left(
\beta >x\right) \sim C\cdot z_{c}^{-x}$ has geometric tails. As detailed
below, 
\begin{equation*}
\phi _{\Delta }\left( z\right) \sim C^{\prime }\cdot \left(
1-z/z_{c}^{\prime }\right) ^{-1}\text{ as }z\rightarrow z_{c}^{\prime }>1
\end{equation*}
so that, with $C^{\prime }=pqC/\left( 1-pC\right) <1$, $\mathbf{P}\left(
\Delta >x\right) \sim C^{\prime }\cdot z_{c}^{\prime -x}$ also has geometric
(heavier) tails but with modified rate $z_{c}^{\prime }=z_{c}\left(
1-pC\right) <z_{c}.$

Then 
\begin{equation*}
\Phi _{\infty }\left( z\right) \sim C^{\prime \prime }\cdot \left(
1-z/z_{c}^{\prime }\right) ^{-1}\text{ as }z\rightarrow z_{c}^{\prime }>1
\end{equation*}
so that $\mathbf{P}\left( X_{\infty }>x\right) \sim C^{\prime }\cdot
z_{c}^{\prime -x}$ also has geometric tails with rate $z_{c}^{\prime }$.
Indeed, with $z_{c}^{\prime \prime }=\left( z_{c}^{\prime }-\left(
1-c\right) \right) /c>z_{c}^{\prime },$%
\begin{eqnarray*}
C^{\prime \prime }\cdot \left( 1-z/z_{c}^{\prime }\right) ^{-1} &\sim
&C^{\prime }C^{\prime \prime }\cdot \left( 1-z/z_{c}^{\prime }\right)
^{-1}\left( 1-\left( 1-c\left( 1-z\right) \right) /z_{c}^{\prime }\right)
^{-1} \\
&=&C^{\prime }C^{\prime \prime }\left( \frac{z_{c}^{\prime }}{cz_{c}^{\prime
\prime }}\right) \cdot \left( 1-z/z_{c}^{\prime }\right) ^{-1}\left(
1-z/z_{c}^{\prime \prime }\right) ^{-1}
\end{eqnarray*}
showing that 
\begin{equation*}
\Phi _{\infty }\left( z\right) \sim C^{\prime }C^{\prime \prime }\left( 
\frac{z_{c}^{\prime }}{cz_{c}^{\prime \prime }}\right) \left(
1-z_{c}^{\prime }/z_{c}^{\prime \prime }\right) ^{-1}\cdot \left(
1-z/z_{c}^{\prime }\right) ^{-1},\text{ as }z\rightarrow z_{c}^{\prime }>1.
\end{equation*}

- Consider a positive recurrent case with $B^{\prime }\left( 1\right) =%
\mathbf{E}\beta =\infty $. This is the case if, with $\alpha \in \left(
0,1\right) $, $B\left( z\right) \sim 1-\left( 1-z\right) ^{\alpha }$ as $%
z\rightarrow 1,$ or if (unscaled Sibuya, see \cite{Sibu}) $B\left( z\right)
=1-\left( 1-z\right) ^{\alpha }$. In this case $\phi _{\Delta }\left(
z\right) \sim 1-\frac{p}{q}\left( 1-z\right) ^{\alpha }$ as a scaled Sibuya
rv (with scale factor $\frac{p}{q}=\mathbf{E}\left( G-1\right) $). Indeed, 
\begin{equation*}
\frac{q}{1-p\left( 1-\left( 1-z\right) ^{\alpha }\right) }=\frac{1}{1+\frac{p%
}{q}\left( 1-z\right) ^{\alpha }}\underset{z\rightarrow 1}{\sim }1-\frac{p}{q%
}\left( 1-z\right) ^{\alpha }
\end{equation*}
Then, $X_{n}$ being recurrent, in view of $\Phi _{\infty }\left( z\right) =%
\frac{q}{1-pB\left( z\right) }\Phi _{\infty }\left( 1-c\left( 1-z\right)
\right) ,$%
\begin{equation*}
\Phi _{\infty }\left( z\right) \underset{z\rightarrow 1}{\sim }1-\gamma
\left( 1-z\right) ^{\alpha }
\end{equation*}
where $\gamma =p/\left[ q\left( 1-c\right) ^{\alpha }\right] $. Indeed, as $%
z\rightarrow 1,$%
\begin{equation*}
1-\gamma \left( 1-z\right) ^{\alpha }\sim \left[ 1-\frac{p}{q}\left(
1-z\right) ^{\alpha }\right] \left[ 1-\gamma c^{\alpha }\left( 1-z\right)
^{\alpha }\right]
\end{equation*}
allowing to identify the scale parameter $\gamma .$ The three rvs $\beta ,$ $%
\Delta $ and $X_{\infty }$ have power law tails with index $\alpha .$

$\bullet $ \textbf{The case }$c=0$\textbf{\ (total disasters)}$.$

In that case, combined to $\Phi _{\infty }\left( 1\right) =1,$%
\begin{equation*}
\Phi _{\infty }\left( z\right) =\frac{q}{1-pB\left( z\right) }=\phi _{\Delta
}\left( z\right)
\end{equation*}
is an admissible pgf solution. The law of $X_{\infty }$ is a compound
shifted-geometric of the $\beta $s, whatever the distribution of $\beta $.

The probabilities $\pi \left( x\right) =\left[ z^{x}\right] \Phi _{\infty
}\left( z\right) $, $x\geq 1$ are explicitly given by the Faa di Bruno
formula for compositions of two pgfs, \cite{Comtet}. Let us look at the
tails of $X_{\infty }$.

- If in particular, with $C\in \left( 0,1\right) $, $B\left( z\right) \sim
C\cdot \left( 1-z/z_{c}\right) ^{-1}$ as $z\rightarrow z_{c}>1$ so that $%
\mathbf{P}\left( \beta >x\right) \sim C\cdot z_{c}^{-x}$ has geometric
tails, then 
\begin{equation*}
\Phi _{\infty }\left( z\right) \sim C^{\prime }\cdot \left(
1-z/z_{c}^{\prime }\right) ^{-1}\text{ as }z\rightarrow z_{c}^{\prime }>1
\end{equation*}
so that, with $C^{\prime }=pqC/\left( 1-pC\right) <1$, $\mathbf{P}\left(
X_{\infty }>x\right) \sim C^{\prime }\cdot z_{c}^{\prime -x}$ also has
geometric (heavier) tails but with modified rate $z_{c}^{\prime
}=z_{c}\left( 1-pC\right) <z_{c}.$

- If $B\left( z\right) =\left( e^{\theta z}-1\right) /\left( e^{\theta
}-1\right) $ ($\beta $ is Poisson conditioned to be positive) is entire. $%
\Phi _{\infty }\left( z\right) $ has a simple pole at $z_{c}>1$ defined by 
\begin{equation*}
\left( e^{\theta z_{c}}-1\right) /\left( e^{\theta }-1\right) =1/p
\end{equation*}
and $X_{\infty }$ has geometric tails with rate $z_{c}$.

- If, with $\alpha \in \left( 0,1\right) $, $B\left( z\right) \sim 1-\left(
1-z\right) ^{\alpha }$ as $z\rightarrow 1,$ or if (unscaled Sibuya) $B\left(
z\right) =1-\left( 1-z\right) ^{\alpha }$, then $\Phi _{\infty }\left(
z\right) \sim 1-\frac{p}{q}\left( 1-z\right) ^{\alpha }$ scaled Sibuya (with
scale factor $\frac{p}{q}=\mathbf{E}\left( G-1\right) :$%
\begin{equation*}
\frac{q}{1-p\left( 1-\left( 1-z\right) ^{\alpha }\right) }=\frac{1}{1+\frac{p%
}{q}\left( 1-z\right) ^{\alpha }}\underset{z\rightarrow 1}{\sim }1-\frac{p}{q%
}\left( 1-z\right) ^{\alpha }
\end{equation*}

- Suppose, with $z_{0}>1$, 
\begin{equation*}
B\left( z\right) =\frac{1-\left( 1-z/z_{0}\right) ^{\alpha }}{1-\left(
1-1/z_{0}\right) ^{\alpha }}
\end{equation*}
If there exists $z_{c}>1:B\left( z_{c}\right) =1/p$ else if $B\left(
z_{0}\right) =1/\left( 1-\left( 1-1/z_{0}\right) ^{\alpha }\right) >1/p$ ($%
\left( 1-1/z_{0}\right) ^{\alpha }>q$ or $z_{0}>1/\left( 1-q^{1/\alpha
}\right) $), then $\Phi _{\infty }\left( z\right) $ has a simple algebraic
pole at $z_{c}.$ If no such $z_{c}$ exists, $\Phi _{\infty }\left( z\right) $
is entire. This is reminiscent of a condensation phenomenon. We finally
observe that

\begin{equation*}
\frac{1-\phi _{\Delta }\left( z\right) }{1-z}=\frac{p\left( 1-B\left(
z\right) \right) }{\left( 1-z\right) \left( 1-pB\left( z\right) \right) }=%
\frac{p}{q}\frac{1-B\left( z\right) }{1-z}\phi _{\Delta }\left( z\right) ,
\end{equation*}
\begin{eqnarray*}
\mathbf{P}\left( \Delta >n\right) &=&\frac{p}{q}\sum_{m=0}^{n}\mathbf{P}%
\left( \beta >n-m\right) \mathbf{P}\left( \Delta =m\right) , \\
\mathbf{P}\left( \Delta =n\right) &=&\frac{p}{q}\left( \sum_{m=0}^{n-1}%
\mathbf{P}\left( \beta =n-m\right) \mathbf{P}\left( \Delta =m\right) +%
\mathbf{P}\left( \Delta =n\right) \right) .
\end{eqnarray*}

When is $\Delta $ with $\phi _{\Delta }\left( z\right) =\frac{q}{1-pB\left(
z\right) }$ itself SD? In any case, $\Delta $ is at least infinitely
divisible (ID) else compound Poisson because $\phi _{\Delta }\left( z\right)
=\exp -r\left( 1-\psi \left( z\right) \right) $ where $r>0$ and $\psi \left(
z\right) $ is a pgf with $\psi \left( 0\right) =0$. Indeed, with $q=e^{-r}$, 
\begin{equation*}
\psi \left( z\right) =\frac{-\log \left( 1-pB\left( z\right) \right) }{-\log
q}
\end{equation*}
is a pgf (the one of a Fisher-log-series rv).\newline

\textbf{Proposition.} \emph{With }$b_{x}=\left[ z^{x}\right] B\left(
z\right) $\emph{, }$x\geq 1$\emph{, the condition } 
\begin{equation}
\frac{b_{x+1}}{b_{x}}\leq \frac{x-pb_{1}}{x+1}\text{ for any }x\geq 1,
\label{CSD}
\end{equation}
\emph{entails that }$\Delta $\emph{\ is SD.}

\emph{Proof:} If $\Delta $ is SD then (see \cite{Sch}, Lemma $2.13$) 
\begin{equation*}
\phi _{\Delta }\left( z\right) =e^{-r\int_{z}^{1}\frac{1-h\left( z^{\prime
}\right) }{1-z^{\prime }}dz^{\prime }},
\end{equation*}
for some $r>0$ and some pgf $h\left( z\right) $ obeying $h\left( 0\right)
=0. $ We are led to check if 
\begin{equation*}
\frac{pB^{\prime }\left( z\right) }{1-pB\left( z\right) }=r\frac{1-h\left(
z\right) }{1-z},
\end{equation*}
for some pgf $h$ and $r=pb_{1}$ where 
\begin{equation*}
h\left( z\right) =1-\frac{1}{b_{1}}\left( 1-z\right) \frac{B^{\prime }\left(
z\right) }{1-pB\left( z\right) }=\frac{1}{b_{1}}\frac{b_{1}\left( 1-pB\left(
z\right) \right) -\left( 1-z\right) B^{\prime }\left( z\right) }{1-pB\left(
z\right) }.
\end{equation*}
Denoting the numerator $N\left( z\right) $, a sufficient condition is that 
\begin{equation*}
\left[ z^{x}\right] N\left( z\right) \geq 0\text{ for all }x\geq 1
\end{equation*}
But 
\begin{equation*}
N\left( z\right) =\sum_{x\geq 1}z^{x}\left[ \left( x-pb_{1}\right)
b_{x}-\left( x+1\right) b_{x+1}\right] .\text{ }\Box
\end{equation*}
Let us show on four examples that these conditions can be met.

1. Suppose $B\left( z\right) =z$. Then $0=\frac{b_{x+1}}{b_{x}}\leq \frac{x-p%
}{x+1}$ for all $x\geq 1$. The simple shifted-geometric rv $\Delta $ is SD.

2. Suppose $B\left( z\right) =b_{1}z+b_{2}z^{2}$ with $b_{2}=1-b_{1}$. We
need to check conditions under which $\frac{b_{2}}{b_{1}}\leq \frac{1-pb_{1}%
}{2}$. This condition is met if and only if the polynomial $%
pb_{1}^{2}-3b_{1}+2\leq 0$ which holds if and only if $b_{1}\geq b_{1}^{*}$
where $b_{1}^{*}\in \left( 0,1\right) $ is the zero of this polynomial in $%
\left( 0,1\right) $.

3. Suppose $B\left( z\right) =\overline{\alpha }z/\left( 1-\alpha z\right) $%
, $\alpha \in \left( 0,1\right) $, the pgf of a geometric$\left( \overline{%
\alpha }\right) $ rv, with $b_{x}=\overline{\alpha }\alpha ^{x-1}.$ The
condition reads: $\alpha \leq \frac{x-p\overline{\alpha }}{x+1}$. It is
fulfilled if $\alpha \leq \frac{x-p}{x+q}$ for all $x\geq 1$ which is $%
\alpha \leq q/\left( 1+q\right) <1$ (or $\overline{\alpha }=b_{1}\geq
b_{1}^{*}=1/\left( 1+q\right) $).

4. Sibuya. Suppose $B\left( z\right) =1-\left( 1-z\right) ^{\alpha }$, $%
\alpha \in \left( 0,1\right) $, with $b_{x}=\alpha \left[ \overline{\alpha }%
\right] _{x-1}/x!$, $x\geq 1$ (where $\left[ \overline{\alpha }\right] _{x}=%
\overline{\alpha }\left( \overline{\alpha }+1\right) ...\left( \overline{%
\alpha }+x-1\right) $, $x\geq 1$ are the rising factorials of $\overline{%
\alpha }$ and $\left[ \overline{\alpha }\right] _{0}:=1$). The condition
reads: $\frac{\overline{\alpha }+x-1}{x+1}\leq \frac{x-p\alpha }{x+1}$ which
is always fulfilled. The shifted-geometric rv with Sibuya distributed
compounding rv is always SD.\newline

\textbf{Proposition}. \emph{Under the condition that }$X_{\infty }$\emph{\
is SD and so unimodal, }$X_{\infty }$\emph{\ has always mode at the origin.}

\emph{Proof:} The condition is that, with $\pi \left( 0\right) =\Phi
_{\infty }\left( 0\right) =q$ and $\pi \left( 1\right) =\Phi _{\infty
}^{\prime }\left( 0\right) =pqB^{\prime }\left( 0\right) =pqb_{1}$, $\pi
\left( 1\right) /\pi \left( 0\right) =pb_{1}<1$ which is always satisfied.

\section{Green (potential) kernel analysis}

In this Section, we analyze some important characteristics of the chain
making use of its Green (potential) kernel.

\subsection{Green (potential) kernel at $\left( 0,0\right) :$ Contact
probability at $0$ and first return time to $0$}

Suppose $X_{0}=0$. With then $\Phi _{0}\left( z\right) =1$, define the
double generating function $\Phi \left( u,z\right) =\sum_{n\geq 0}u^{n}\Phi
_{n}\left( z\right) $, obeying $\Phi \left( u,1\right) =1/\left( 1-u\right) $%
, see \cite{Woess}. [Note that $\log z$ was Laplace-conjugate to $X_{n}$ and
now $\log u$ is Laplace-conjugate to $n$]. Then, 
\begin{equation*}
\frac{1}{u}\left( \Phi \left( u,z\right) -1\right) =pB\left( z\right) \Phi
\left( u,z\right) +q\Phi \left( u,1-c\left( 1-z\right) \right)
\end{equation*}
\begin{equation}
\Phi \left( u,z\right) =\frac{1+qu\Phi \left( u,1-c\left( 1-z\right) \right) 
}{1-puB\left( z\right) }  \label{double}
\end{equation}
\begin{equation*}
\Phi \left( u,0\right) =1+qu\Phi \left( u,1-c\right) .
\end{equation*}

With $H\left( u,z\right) :=1/\left( 1-puB\left( z\right) \right) ,$ upon
iterating, with $\Phi \left( 0,z\right) =\Phi _{0}\left( z\right) =1$%
\begin{equation}
\Phi \left( u,z\right) =\sum_{n\geq 0}\left( qu\right)
^{n}\prod_{m=0}^{n}H\left( u,1-c^{m}\left( 1-z\right) \right) .  \label{it}
\end{equation}
In particular 
\begin{equation}
\Phi \left( u,0\right) =\sum_{n\geq 0}\left( qu\right)
^{n}\prod_{m=0}^{n}H\left( u,1-c^{m}\right) =1+\sum_{n\geq 1}\prod_{m=1}^{n}%
\frac{qu}{1-puB\left( 1-c^{m}\right) }.  \label{G00}
\end{equation}
Note that 
\begin{equation*}
G_{0,0}\left( u\right) :=\Phi \left( u,0\right) =\sum_{n\geq 0}u^{n}\mathbf{P%
}_{0}\left( X_{n}=0\right) =\left( I-uP\right) ^{-1}\left( 0,0\right)
\end{equation*}
is the Green kernel of the chain at $\left( 0,0\right) $ (the matrix element 
$\left( 0,0\right) $ of the resolvent of $P$). Consequently, \newline

\textbf{Proposition.} \emph{The Green kernel }$G_{0,0}\left( u\right) $\emph{%
\ is given by (\ref{G00}). }

\emph{With }$h_{m^{\prime }}:=\left[ u^{m^{\prime }}\right]
\prod_{m=0}^{n-1}\left( 1-puB\left( 1-c^{m}\right) \right) ^{-1}$\emph{, we
have the following expression for the contact probability at }$0:$%
\begin{equation}
\left[ u^{n}\right] \Phi \left( u,0\right) =\Phi _{n}\left( 0\right) =%
\mathbf{P}_{0}\left( X_{n}=0\right) =\sum_{m^{\prime
}=0}^{n-1}q^{n-m^{\prime }}h_{m^{\prime }}\text{.}  \label{cont}
\end{equation}

\textbf{Remarks.}

$\left( i\right) $\textit{\ Let us have a quick check of this formula. When }%
$n=1$\textit{, this leads to }$\mathbf{P}_{0}\left( X_{1}=0\right) =q,$%
\textit{\ and, if }$n=2$\textit{\ to }$\mathbf{P}_{0}\left( X_{2}=0\right)
=q^{2}+q\left[ u^{1}\right] \left( 1-puB\left( 1-c\right) \right)
^{-1}=q^{2}+pqB\left( 1-c\right) .$\textit{\ The second part is not quite
trivial because it accounts for any movement up in the first step (wp }$p$%
\textit{) immediately followed by a movement down to }$0$\textit{. This is
consistent however with the binomial formula }$\mathbf{P}\left( X_{1}=0\mid
X_{0}=x\right) =q\mathbf{P}\left( c\circ x=0\right) =q\left( 1-c\left(
1-z\right) \right) ^{x}\mid _{z=0}=q\left( 1-c\right) ^{x}$\textit{\ so that
the second part is}

\begin{equation*}
p\sum_{x\geq 1}\mathbf{P}\left( \beta =x\right) \mathbf{P}\left( X_{1}=0\mid
X_{0}=x\right) =pq\sum_{x\geq 1}b_{x}\left( 1-c\right) ^{x}=pqB\left(
1-c\right) .
\end{equation*}
\textit{\newline
}

$\left( ii\right) $\textit{\ With }$B_{m}:=pB\left( 1-c^{m}\right) $\textit{%
, }$m\geq 1,$\textit{\ an increasing }$\left[ 0,1\right] -$\textit{\ valued
sequence converging to }$p$\textit{, decomposing the product }$%
\prod_{m=1}^{n-1}\left( 1-uB_{m}\right) ^{-1}$\textit{\ into simple fraction
elements, we get} \textit{if} $n>1$%
\begin{equation*}
h_{m^{\prime }}=\sum_{m=1}^{n-1}A_{m,n}B_{m}^{m^{\prime }},
\end{equation*}
\textit{where }$A_{m,n}=B_{m}^{n-2}/\prod_{m^{\prime }\in \left\{
1,..,n-1\right\} \backslash \left\{ m\right\} }\left( B_{m}-B_{m^{\prime
}}\right) .$\textit{\ The term }$A_{n-1}B_{n-1}^{m^{\prime }}$\textit{\
contributes most to the sum. Hence, }$\mathbf{P}_{0}\left( X_{1}=0\right) =q$%
\textit{\ and if }$n>1$%
\begin{eqnarray*}
\mathbf{P}_{0}\left( X_{n}=0\right) &=&q^{n}\sum_{m^{\prime
}=0}^{n-1}q^{-m^{\prime }}\sum_{m=1}^{n-1}A_{m,n}B_{m}^{m^{\prime
}}=q^{n}\sum_{m=1}^{n-1}A_{m,n}\sum_{m^{\prime }=0}^{n-1}\left( \frac{B_{m}}{%
q}\right) ^{m^{\prime }} \\
&=&q\sum_{m=1}^{n-1}A_{m,n}\frac{q^{n}-B_{m}^{n}}{q-B_{m}}
\end{eqnarray*}
\textit{is an alternative representation of (\ref{cont})}.

\textit{In the positive recurrent case, as }$n\rightarrow \infty ,$ 
\begin{equation*}
\mathbf{P}_{0}\left( X_{n}=0\right) \rightarrow \pi \left( 0\right)
=\prod_{n\geq 0}\phi _{\Delta }\left( 1-c^{n}\right) >0.
\end{equation*}
\newline

The Green kernel at $\left( 0,0\right) $ is thus $G_{0,0}\left( u\right)
=\Phi \left( u,0\right) $.

If $n\geq 1,$ from the recurrence $\mathbf{P}_{0}\left( X_{n}=0\right)
=P^{n}\left( 0,0\right) =\sum_{m=0}^{n}\mathbf{P}\left( \tau _{0,0}=m\right)
P^{n-m}\left( 0,0\right) $, we see that the pgf $\phi _{0,0}\left( u\right) =%
\mathbf{E}\left( u^{\tau _{0,0}}\right) $ of the first return time to $0$, $%
\tau _{0,0}$ and $G_{0,0}\left( u\right) $ are related by the Feller
relation (see \cite{Bing} pp $3-4$ for example) 
\begin{equation*}
G_{0,0}\left( u\right) =\frac{1}{1-\phi _{0,0}\left( u\right) }\text{ and }%
\phi _{0,0}\left( u\right) =\frac{G_{0,0}\left( u\right) -1}{G_{0,0}\left(
u\right) }.
\end{equation*}
Hence,

\textbf{Proposition.} \emph{The pgf }$\phi _{0,0}\left( u\right) $\emph{\ of
the first return time }$\tau _{0,0}$\emph{\ is } 
\begin{equation}
\phi _{0,0}\left( u\right) =1-\frac{1}{G_{0,0}\left( u\right) },
\label{Tau00}
\end{equation}
\emph{where }$G_{0,0}\left( u\right) $\emph{\ is given by (\ref{G00}).}

Note 
\begin{equation*}
G_{0,0}\left( 1\right) =\sum_{n\geq 0}\mathbf{P}_{0}\left( X_{n}=0\right)
=1+\sum_{n\geq 1}q^{n}\prod_{m=0}^{n}H\left( 1,1-c^{m}\right) =\infty
\end{equation*}
if and only if $X$ is recurrent, \cite{Neveu}, \cite{Sato}. And in that
case, $\phi _{0,0}\left( 1\right) =\mathbf{P}\left( \tau _{0,0}<\infty
\right) =1-\frac{1}{G_{0,0}\left( 1\right) }=1.$ Positive-(- null)
recurrence is when $\phi _{0,0}^{\prime }\left( 1\right) =\mathbf{E}\left(
\tau _{0,0}\right) =1/\pi _{0}<\infty $ ($=\infty $), \cite{Kac}. Note
finally $G_{0,0}\left( 0\right) =1$ so that $\phi _{0,0}\left( 0\right) =%
\mathbf{P}\left( \tau _{0,0}^{{}}=0\right) =0$.

\subsection{Starting from $x_{0}>0:$ Green kernel at $\left( x_{0},0\right) $
and first extinction time $\tau _{x_{0},0}$}

Suppose now\textbf{\ }$X_{0}=x_{0}>0.$ After shifting $X_{n}$ of $x_{0}$,
with $\Phi \left( u,z\right) =\sum_{n\geq 0}u^{n}\mathbf{E}\left(
z^{x_{0}+X_{n}}\right) $, we now get 
\begin{equation*}
\frac{1}{u}\left( \Phi \left( u,z\right) -z^{x_{0}}\right) =pB\left(
z\right) \Phi \left( u,z\right) +q\Phi \left( u,1-c\left( 1-z\right) \right)
.
\end{equation*}
Then 
\begin{equation}
\Phi \left( u,z\right) =\frac{z^{x_{0}}+qu\Phi \left( u,1-c\left( 1-z\right)
\right) }{1-puB\left( z\right) },  \label{it2}
\end{equation}
entailing 
\begin{eqnarray*}
\Phi \left( u,z\right) &=&\sum_{n\geq 0}\left( qu\right) ^{n}\left(
1-c^{n}\left( 1-z\right) \right) ^{x_{0}}\prod_{m=0}^{n}H\left(
u,1-c^{m}\left( 1-z\right) \right) , \\
\Phi \left( u,0\right) &=&\sum_{n\geq 0}\left( qu\right) ^{n}\left(
1-c^{n}\right) ^{x_{0}}\prod_{m=0}^{n}H\left( u,1-c^{m}\right) .
\end{eqnarray*}
We obtained:

\textbf{Proposition.} \emph{The contact probability at }$0$\emph{\ for the
chain started at }$x_{0}>0$ \emph{is given by} 
\begin{equation}
\left[ u^{n}\right] \Phi \left( u,0\right) =\Phi _{n}\left( 0\right) =%
\mathbf{P}_{x_{0}}\left( X_{n}=0\right) =\sum_{m^{\prime }=0}^{n-1}\left(
1-c^{n-m^{\prime }}\right) ^{x_{0}}q^{n-m^{\prime }}h_{m^{\prime }}.
\label{cont2}
\end{equation}

Let us give the first two terms as compared to when $x_{0}=0$. As required,
when $n=1$, $\mathbf{P}_{x_{0}}\left( X_{1}=0\right) =q\left( 1-c\right)
^{x_{0}}$ and when $n=2$, 
\begin{equation*}
\mathbf{P}_{x_{0}}\left( X_{2}=0\right) =q^{2}\left( 1-c^{2}\right)
^{x_{0}}+pq\left( 1-c\right) ^{x_{0}}B\left( 1-c\right)
\end{equation*}
a weighted sum of the two terms appearing in the above expression of $%
\mathbf{P}_{0}\left( X_{2}=0\right) .$\newline

\textbf{Corollary.}

$\left( i\right) $\emph{\ When }$x_{0}$\emph{\ is large and }$n$\emph{\
fixed, the small but dominant term is when }$m^{\prime }=0$\emph{\ which is }%
$q^{n}\left( 1-c^{n}\right) ^{x_{0}}.$\emph{\ So }$\mathbf{P}_{x_{0}}\left(
X_{n}=0\right) $\emph{\ decays geometrically with }$x_{0}$\emph{. This
expression quantifies the probability that the population is in an early
state of extinction given the initial population size was large. Early is
when }$c^{n}\gg 1/x_{0}$\emph{\ (so that }$\left( 1-c^{n}\right) ^{x_{0}}\ll
\left( 1-1/x_{0}\right) ^{x_{0}}\ll e^{-1}$\emph{), so when }$n\ll -\log
_{c}x_{0}.$

$\left( ii\right) $\emph{\ In the transient case, when }$n$\emph{\ is large
and }$x_{0}$\emph{\ is fixed, the dominant term is when }$m^{\prime }=n-1$%
\emph{\ which is }$q\left( 1-c\right) ^{x_{0}}\left( pB\left( 1-c\right)
\right) ^{n}.$\emph{\ So }$\mathbf{P}_{x_{0}}\left( X_{n}=0\right) $\emph{\
decays geometrically with }$n$\emph{\ at rate }$pB\left( 1-c\right) $\emph{.
In the positive recurrent case, }$\mathbf{P}_{x_{0}}\left( X_{n}=0\right)
\rightarrow \pi \left( 0\right) >0$\emph{\ as }$n\rightarrow \infty $\emph{,
independently of }$x_{0}$\emph{.}

$\bullet $ \textbf{The} \textbf{Green kernel at }$\left( x_{0},0\right) $ is
thus $G_{x_{0},0}\left( u\right) =\left[ z^{0}\right] \Phi \left( u,z\right) 
$ (the matrix element $\left( x_{0},0\right) $ of the resolvent of $P$)$.$
It is related to the pgf of the first extinction time $\tau _{x_{0},0}$ by
the Feller relation 
\begin{equation}
\phi _{x_{0},0}\left( u\right) =\mathbf{E}\left( u^{\tau _{x_{0},0}}\right) =%
\frac{G_{x_{0},0}\left( u\right) }{G_{0,0}\left( u\right) }.  \label{phix00}
\end{equation}
Therefore,

\textbf{Proposition.} \emph{With }$x_{0}>0$\emph{, the pgf of the first
extinction time }$\tau _{x_{0},0}$\emph{\ is } 
\begin{equation}
\phi _{x_{0},0}\left( u\right) =\mathbf{E}\left( u^{\tau _{x_{0},0}}\right) =%
\frac{\sum_{n\geq 1}\left( qu\right) ^{n}\left( 1-c^{n}\right)
^{x_{0}}\prod_{m=1}^{n}H\left( u,1-c^{m}\right) }{\sum_{n\geq 0}\left(
qu\right) ^{n}\prod_{m=0}^{n}H\left( u,1-c^{m}\right) }.  \label{taux00}
\end{equation}

In the recurrent case, state $0$ is visited infinitely often and so both $%
G_{0,0}\left( 1\right) $ and $G_{x_{0},0}\left( 1\right) =\infty $, and 
\begin{equation*}
\mathbf{P}\left( \tau _{x_{0},0}<\infty \right) =\phi _{x_{0},0}\left(
1\right) =\frac{G_{x_{0},0}\left( 1\right) }{G_{0,0}\left( 1\right) }=1.
\end{equation*}
We finally note that, because state $0$ is reflecting, $\tau _{x_{0},0}$ is
only a local extinction time, followed by subsequent extinction times after $%
\tau _{0,0}.$ In the sequel, we shall let $\overline{P}$ stand for the
substochastic transition matrix obtained from $P$ while deleting its first
row and column.\newline

There are two alternative representations of $\phi _{x_{0},0}\left( u\right)
.$

$\bullet $ One alternative representation follows from the following
classical first-step analysis:

Let $X_{1}\left( x_{0}\right) $ be the position of $X_{n}$ started at $x_{0}$%
.

Let $X_{+}\left( x_{0}\right) $ be a positive rv with $\mathbf{P}\left(
X_{+}\left( x_{0}\right) =y\right) =\overline{P}\left( x_{0},y\right)
/\sum_{y\geq 1}\overline{P}\left( x_{0},y\right) $, $y\geq 1$. With $\tau
_{X_{+}\left( x_{0}\right) ,0}^{\prime }$ a statistical copy of $\tau
_{X_{+}\left( x_{0}\right) ,0}$, first-step analysis yields, (see \cite{Nor}%
, \cite{Woess}): 
\begin{equation*}
\tau _{x_{0},0}\overset{d}{=}1\cdot \mathbf{1}_{\left\{ X_{1}\left(
x_{0}\right) =0\right\} }+\mathbf{1}_{\left\{ X_{1}\left( x_{0}\right)
>0\right\} }\cdot \left( 1+\tau _{X_{+}\left( x_{0}\right) ,0}^{\prime
}\right) .
\end{equation*}
Clearly, $\mathbf{P}\left( X_{1}\left( x_{0}\right) =0\right) =P\left(
x_{0},0\right) =q\left( 1-c\right) ^{x_{0}}=:q_{x_{0}}$, $\mathbf{P}\left(
X_{1}\left( x_{0}\right) >0\right) =:p_{x_{0}}=\sum_{y\geq 1}\overline{P}%
\left( x_{0},y\right) =1-q_{x_{0}}.$ Therefore $\phi _{x_{0},0}\left(
u\right) :=\mathbf{E}\left( u^{\tau _{x_{0},0}}\right) ,$ obeys the
recurrence $\phi _{x_{0},0}\left( u\right) =q_{x_{0}}u+up_{x_{0}}\mathbf{E}%
\phi _{X_{+}\left( x_{0}\right) ,0}\left( u\right) $.

With $\mathbf{\phi }\left( u\right) =\left( \phi _{1,0}\left( u\right) ,\phi
_{2,0}\left( u\right) ,...\right) ^{\prime }$ the column-vector of the $\phi
_{x_{0},0}\left( u\right) =\mathbf{E}u^{\tau _{x_{0},0}}$, and $\mathbf{q}%
=\left( q_{1},q_{2},...\right) ^{\prime }$ the first column-vector of the
matrix $P,$ $\mathbf{\phi }\left( u\right) $ then solves: 
\begin{equation}
\mathbf{\phi }\left( u\right) =u\mathbf{q}+u\overline{P}\mathbf{\phi }\left(
u\right) ,  \label{4}
\end{equation}
whose formal solution is (compare with the explicit expression (\ref{taux00}%
)) 
\begin{equation}
\mathbf{\phi }\left( u\right) =u\left( I-u\overline{P}\right) ^{-1}\mathbf{q}%
=:u\overline{G}\left( u\right) \mathbf{q,}  \label{R1}
\end{equation}
involving the resolvent matrix $\overline{G}\left( u\right) $ of $\overline{P%
}.$ Note $\mathbf{\phi }\left( 1\right) =\left( I-\overline{P}\right) ^{-1}%
\mathbf{q}$ gives the column-vector of the probabilities of eventual
extinction $\mathbf{\phi }\left( 1\right) :=\left( \phi _{1,0}\left(
1\right) ,\phi _{2,0}\left( 1\right) ,...\right) ^{\prime }$, so with $\phi
_{x_{0},0}\left( 1\right) =\mathbf{P}\left( \tau _{x_{0},0}<\infty \right) $
if $\overline{G}\left( 1\right) \mathbf{q}<\infty .$ Clearly $\mathbf{\phi }%
\left( 1\right) =\mathbf{1}$ (the all-one column vector) in the recurrent
case. In that case, from (\ref{4}), introducing the column vector $\mathbf{E}%
\left( \tau _{.,0}\right) :=\left( \mathbf{E}\left( \tau _{1,0}\right) ,%
\mathbf{E}\left( \tau _{2,0}\right) ,...\right) ^{\prime }$ where $\mathbf{E}%
\left( \tau _{x_{0},0}\right) =\phi _{x_{0},0}^{\prime }\left( 1\right) $
and observing $\mathbf{q}+\overline{P}\mathbf{\phi }\left( 1\right) =\mathbf{%
1}$, we get 
\begin{equation*}
\mathbf{\phi }^{\prime }\left( 1\right) :=\mathbf{E}\left( \tau
_{.,0}\right) =\mathbf{1}+\overline{P}\mathbf{E}\left( \tau _{.,0}\right) ,%
\text{ equivalently }\mathbf{E}\left( \tau _{.,0}\right) =\left( I-\overline{%
P}\right) ^{-1}\mathbf{1}=\overline{G}\left( 1\right) \mathbf{1,}\text{ or}
\end{equation*}
\begin{equation*}
\mathbf{E}\left( \tau _{x_{0},0}\right) =\sum_{y\geq 1}\overline{G}%
_{x_{0},y}\left( 1\right) \mathbf{,}
\end{equation*}
where $\overline{G}_{x_{0},y}$ is the matrix element $\left( x_{0},y\right) $
of the resolvent of $\overline{P}$.\newline

$\bullet $ Yet another alternative representation $\mathbf{\phi }\left(
u\right) :=\left( \phi _{1,0}\left( u\right) ,\phi _{2,0}\left( u\right)
,...\right) ^{\prime }$ is as follows. From the identity 
\begin{equation*}
\overline{P}^{n}\left( x_{0},y\right) =\mathbf{P}_{x_{0}}\left( X_{n}=y,\tau
_{x_{0},0}>n\right) ,
\end{equation*}
we get $\mathbf{P}\left( \tau _{.,0}>n\right) =\overline{P}^{n}\mathbf{1},$%
as the column vector of $\left( \mathbf{P}\left( \tau _{1,0}>n\right) ,\text{
}\mathbf{P}\left( \tau _{2,0}>n\right) ,...\right) ^{\prime }$, and so, 
\begin{equation*}
\sum_{n\geq 0}u^{n}\mathbf{P}\left( \tau _{.,0}>n\right) =\overline{G}\left(
u\right) \mathbf{1.}
\end{equation*}
This leads in particular, as expected, to $\mathbf{E}\left( \tau
_{.,0}\right) =\overline{G}\left( 1\right) \mathbf{1}$ and (compare with (%
\ref{R1}) and the explicit expression (\ref{taux00})) to 
\begin{equation}
\mathbf{\phi }\left( u\right) =\sum_{n\geq 0}u^{n}\mathbf{P}\left( \tau
_{.,0}=n\right) =\mathbf{1}-\left( 1-u\right) \overline{G}\left( u\right) 
\mathbf{1.}  \label{R2}
\end{equation}
We obtained:

\textbf{Proposition:} \emph{We have }$\mathbf{\phi }\left( 1\right) =\mathbf{%
P}\left( \tau _{.,0}<\infty \right) $\emph{\ and so }$\overline{G}\left(
1\right) \mathbf{1}<\infty \Rightarrow \mathbf{P}\left( \tau _{.,0}<\infty
\right) =\mathbf{1}$\emph{, meaning recurrence of }$X_{n}.$\emph{\ In fact,
positive recurrence is precisely when }$\mathbf{E}\left( \tau _{.,0}\right) =%
\overline{G}\left( 1\right) \mathbf{1}<\infty $\emph{. If }$\overline{G}%
\left( 1\right) \mathbf{1}=\infty ,$ \emph{the chain is null-recurrent if }$%
\left( 1-u\right) \overline{G}\left( u\right) \mathbf{1\rightarrow 0}$ as $%
u\rightarrow 1$, \emph{transient if }$\left( 1-u\right) \overline{G}\left(
u\right) \mathbf{1\rightarrow P}\left( \tau _{.,0}=\infty \right) $\emph{\ as%
} $u\rightarrow 1$\emph{, a non-zero limit.}\newline

The matrix $\overline{P}$ is substochatic with spectral radius $\rho \in
\left( 0,1\right) .$ With $\mathbf{r}$ and $\mathbf{l}^{\prime }$ the
corresponding right and left positive eigenvectors of $\overline{P}$, so
with $\overline{P}\mathbf{r}=\rho \mathbf{r}$ and $\mathbf{l}^{\prime }%
\overline{P}=\rho \mathbf{l}^{\prime }$, $\overline{P}^{n}\sim \rho
^{n}\cdot \mathbf{rl}^{\prime }$ (as $n$ is large) where $\mathbf{rl}%
^{\prime }$ is the projector onto the first eigenspace. By Perron-Frobenius
theorem, \cite{VJ}, \cite{GD}, we can normalize $\mathbf{l}$ to be of $%
l_{1}- $ norm one to get\newline

\textbf{Proposition: }\emph{In the positive recurrent case for }$X_{n}$ ($%
\mathbf{E}\log _{+}\Delta <\infty $)$:$

$\left( i\right) $ \emph{With }$r\left( x_{0}\right) $\emph{\ the }$x_{0}-$%
\emph{entry of }$\mathbf{r}$\emph{,} 
\begin{equation}
\rho ^{-n}\mathbf{P}\left( \tau _{x_{0},0}>n\right) \rightarrow r\left(
x_{0}\right) ,\text{ as }n\rightarrow \infty ,  \label{Edecay}
\end{equation}
\emph{showing that }$\mathbf{P}\left( \tau _{x_{0},0}>n\right) $\emph{\ has
geometric tails with rate }$\rho $ \emph{(extinction is fast)}.

$\left( ii\right) $ \emph{With }$l\left( y\right) $\emph{\ the }$y-$\emph{%
entry of }$\mathbf{l,}$ \emph{for all }$x_{0}>0,$%
\begin{equation}
\mathbf{P}_{x_{0}}\left( X_{n}=y\mid \tau _{x_{0},0}>n\right) \rightarrow
l\left( y\right) ,\text{ as }n\rightarrow \infty ,  \label{qsdl}
\end{equation}
\emph{showing that the left eigenvector }$\mathbf{l}$\emph{\ is the
quasi-stationary distribution of }$X_{n}$\emph{\ (or Yaglom limit, \cite{Ya}%
), \cite{CMSM}.}

\emph{Proof:} In this case, with $R=\rho ^{-1}>1,$ the convergence radius of 
$\overline{G}$, $\overline{G}\left( R\right) =\infty $ and $\overline{P}$ is 
$R-$positive recurrent. $\left( i\right) $ follows\emph{\ }from\emph{\ }$%
\mathbf{P}\left( \tau _{.,0}>n\right) =\overline{P}^{n}\mathbf{1}$ and $%
\left( ii\right) $ from $\mathbf{P}_{x_{0}}\left( X_{n}=y\mid \tau
_{x_{0},0}>n\right) =\overline{P}^{n}\left( x_{0},y\right) /\overline{P}^{n}%
\mathbf{1.}$\newline

\textbf{Remark.} \textit{The full Green kernel at\ }$\left(
x_{0},y_{0}\right) $\textit{\ is }$G_{x_{0},y_{0}}\left( u\right) =\left[
z^{y_{0}}\right] \Phi \left( u,z\right) $\textit{. Hence } 
\begin{equation}
G_{x_{0},y_{0}}\left( u\right) =\sum_{n\geq 0}\left( qu\right) ^{n}\left[
z^{y_{0}}\right] \left( 1-c^{n}\left( 1-z\right) \right)
^{x_{0}}\prod_{m=0}^{n}H\left( u,1-c^{m}\left( 1-z\right) \right)
\label{Green}
\end{equation}
\begin{equation*}
=\sum_{n\geq 0}\left( qu\right) ^{n}\sum_{y=0}^{y_{0}}h_{n,y}\left( u\right)
g_{n,y_{0}-y}
\end{equation*}
\textit{where } 
\begin{equation*}
g_{n,y}=\left[ z^{y}\right] \left( 1-c^{n}\left( 1-z\right) \right) ^{x_{0}}=%
\binom{x_{0}}{y}c^{ny}\left( 1-c^{n}\right) ^{x_{0}-y}
\end{equation*}
\textit{and } 
\begin{equation*}
h_{n,y}\left( u\right) =\left[ z^{y}\right] \prod_{m=0}^{n}H\left(
u,1-c^{m}\left( 1-z\right) \right) ,
\end{equation*}
\textit{which can be obtained from a decomposition into simple elements of
the inner product.}

\textit{Using }$P^{n}\left( x_{0},y_{0}\right) =\sum_{m=1}^{n}\Bbb{P}\left(
\tau _{x_{0},y_{0}}=m\right) P^{n-m}\left( y_{0},y_{0}\right) $\textit{, }$%
n\geq 1$\textit{, we easily get the expression of the pgf of the first
hitting times }$\tau _{x_{0},y_{0}}=\inf \left( n\geq 1:X_{n}=y_{0}\mid
X_{0}=x_{0}\right) $\textit{, as } 
\begin{equation}
\phi _{x_{0},y_{0}}\left( z\right) =\frac{G_{x_{0},y_{0}}\left( z\right) }{%
G_{y_{0},y_{0}}\left( z\right) }.  \label{hit}
\end{equation}

\subsection{\textbf{Average position of }$X_{n}$\textbf{\ started at }$x_{0}$%
\textbf{\ }}

The double-generating function $\Phi \left( u,z\right) $ can be used to
compute the evolution of $\mathbf{E}_{x_{0}}\left( X_{n}\right) $. With%
\textbf{\ }$\Phi ^{\prime }\left( u,z\right) =\partial _{z}\Phi \left(
u,z\right) ,$

\begin{eqnarray*}
\Phi ^{\prime }\left( u,1\right) &=&x_{0}\sum_{n\geq 0}\left( qcu\right)
^{n}H\left( u,1\right) ^{n+1}+\frac{p\rho }{1-\left( p+qc\right) }\left( 
\frac{1}{1-u}-\frac{1}{1-u\left( p+qc\right) }\right) \\
&=&x_{0}\frac{H\left( u,1\right) }{1-qcuH\left( u,1\right) }+\frac{p\rho }{%
1-\left( p+qc\right) }\left( \frac{1}{1-u}-\frac{1}{1-u\left( p+qc\right) }%
\right) \\
&=&\frac{x_{0}}{1-\left( p+qc\right) u}+\frac{p\rho }{1-\left( p+qc\right) }%
\left( \frac{1}{1-u}-\frac{1}{1-u\left( p+qc\right) }\right)
\end{eqnarray*}
This shows that, in the positive recurrent case, with $\mu _{n}=\mathbf{E}%
_{x_{0}}\left( X_{n}\right) ,$%
\begin{equation}
\mu _{n}=\left[ u^{n}\right] \Phi ^{\prime }\left( u,1\right) =x_{0}\left(
p+qc\right) ^{n}+\frac{p\rho }{q\left( 1-c\right) }\left( 1-\left( 1-q\left(
1-c\right) \right) ^{n}\right)  \label{avera}
\end{equation}
\begin{equation*}
=x_{0}\left( p+qc\right) ^{n}+\mathbf{E}_{0}\left( X_{n}\right) \rightarrow 
\frac{p\rho }{q\left( 1-c\right) }\text{ whatever }x_{0},
\end{equation*}
solving 
\begin{equation}
\mu _{n+1}=\left( p+qc\right) \mu _{n}+p\rho ,\text{ }\mu _{0}=x_{0}.
\label{Rmean}
\end{equation}
A similar analysis making use of the second derivative $\Phi ^{^{\prime
\prime }}\left( u,1\right) $ would yield the transient evolution of the
variance of $X_{n},$ started at $x_{0}.$ We skip the details.

\subsection{The height $H$ of an excursion}

Assume $X_{0}=x_{0}$ and consider a version of $X_{n}$ which is absorbed at $%
0.$ Let $X_{n\wedge \tau _{x_{0},0}}^{{}}$ stopping $X_{n}$ when it first
hits $0.$ Let us define the scale (or harmonic) function or sequence $%
\varphi $ of $X_{n}$ as the function which makes $Y_{n}\equiv \varphi \left(
X_{n\wedge \tau _{x_{0},0}}^{{}}\right) $ a martingale. The function $%
\varphi $ is important because, for all $0\leq x_{0}<h,$ with $\tau
_{x_{0}}=\tau _{x_{0},0}\wedge \tau _{x_{0},h}$ the first hitting time of $%
\left\{ 0,h\right\} $ starting from $x_{0}$ (assuming $\varphi \left(
0\right) \equiv 0$)

\begin{equation*}
\mathbf{P}\left( X_{\tau _{x_{0}}}=h\right) =\mathbf{P}\left( \tau
_{x_{0},h}<\tau _{x_{0},0}\right) =\frac{\varphi \left( x_{0}\right) }{%
\varphi \left( h\right) },
\end{equation*}
resulting from 
\begin{equation*}
\mathbf{E}\varphi \left( X_{n\wedge \tau _{x_{0}}}^{{}}\right) =\varphi
\left( x_{0}\right) =\varphi \left( h\right) \mathbf{P}\left( \tau
_{x_{0},h}<\tau _{x_{0},0}\right) +\varphi \left( 0\right) \mathbf{P}\left(
\tau _{x_{0},h}>\tau _{x_{0},0}\right) .
\end{equation*}

$\bullet $ \textbf{The case }$\beta _{1}\sim \delta _{1}$\textbf{.} Let us
consider the height $H$ of an excursion of the original MC $X_{n}$ first
assuming $\beta _{1}\sim \delta _{1}$ (a birth event adds only one
individual)$.$ With probability $q$, $H=0$ and with probability $p$,
starting from $X_{1}=1$, it is the height of a path from state $1$ to $0$ of
the absorbed process $X_{n}$. Using this remark, the event $H=h$ is realized
when $\tau _{1,h}<\tau _{1,0}$ and $\tau _{h,h+1}>\tau _{h,0},$ the latter
two events being independent. Thus (with $\mathbf{P}\left( H=0\right) =q$): 
\begin{equation}
\mathbf{P}\left( H=h\right) =p\frac{\varphi \left( 1\right) }{\varphi \left(
h\right) }\left( 1-\frac{\varphi \left( h\right) }{\varphi \left( h+1\right) 
}\right) \text{, }h\geq 1.  \label{h1}
\end{equation}
We clearly have $\sum_{h\geq 1}\mathbf{P}\left( H=h\right) =p$ because
partial sums form a telescoping series. But (\ref{h1}) is also 
\begin{equation}
\mathbf{P}\left( H\geq h\right) =1/\varphi \left( h\right) ,\text{ }h\geq 1,
\label{h2}
\end{equation}
with $\varphi \left( 1\right) =1/p$.

It remains to compute $\varphi $ with $\varphi \left( 0\right) =0$ and $%
\mathbf{P}\left( H\geq 1\right) =1/\varphi \left( 1\right) =p$. We wish to
have: $\mathbf{E}_{x_{0}}\left( Y_{n+1}\mid Y_{n}=x\right) =x$, leading to 
\begin{equation*}
\varphi \left( x\right) =p\varphi \left( x+1\right) +q\sum_{y=1}^{x}\binom{x%
}{y}c^{y}\left( 1-c\right) ^{x-y}\varphi \left( y\right) ,\text{ }x_{0}\geq
1.
\end{equation*}
The vector $\mathbf{\varphi }$ is the right eigenvector associated to the
eigenvalue $1$ of the modified version $P^{*}$ of the stochastic matrix $P$
having $0$ as an absorbing state: ($P^{*}\left( 0,0\right) =1$), so with: $%
\mathbf{\varphi }=P^{*}\mathbf{\varphi }$, $\varphi \left( 0\right) =0$, 
\cite{Nor}. The searched `harmonic' function is increasing and given by
recurrence, $\varphi \left( 1\right) =1/p$ and 
\begin{equation}
\varphi \left( x+1\right) =\frac{1}{p}\left( \varphi \left( x\right) \left[
1-qc^{x}\right] -q\sum_{y=1}^{x-1}\binom{x}{y}c^{y}\left( 1-c\right)
^{x-y}\varphi \left( y\right) \right) ,\text{ }x\geq 1  \label{h3}
\end{equation}
The first two terms are 
\begin{eqnarray*}
\varphi \left( 2\right) &=&\frac{1}{p}\left( 1-qc\right) \varphi \left(
1\right) =\frac{1}{p^{2}}\left( 1-qc\right) , \\
\varphi \left( 3\right) &=&\frac{1}{p}\left( \varphi \left( 2\right) \left[
1-qc^{2}\right] -2qc\left( 1-c\right) \varphi \left( 1\right) \right) , \\
&=&\frac{1}{p^{3}}\left( 1-qc\right) \left( 1-qc^{2}\right) -2\frac{q}{p}%
c\left( 1-c\right) .
\end{eqnarray*}
The sequence $\varphi \left( x\right) $ is diverging when the chain $X$ is
recurrent. \newline

\textbf{Proposition.} \emph{When }$\beta \sim \delta _{1}$\emph{, equations (%
\ref{h2}) and (\ref{h3}) characterize the law of the excursion height }$H$%
\emph{\ of the random walker in the recurrent case. In the transient case, }$%
\varphi \left( x\right) $\emph{\ converges to a value }$\varphi ^{*}$\emph{\
and }$\mathbf{P}\left( H=\infty \right) =1/\varphi ^{*}=\mathbf{P}\left(
\tau _{0,0}=\infty \right) .$\newline

$\bullet $ \textbf{General }$\beta _{1}$\textbf{.} Whenever the law of $%
\beta $\ is general, the matrix $P^{*}$\ is no longer lower Hessenberg and
the harmonic vector $\mathbf{\varphi }=P^{*}\mathbf{\varphi }$, with $%
\varphi \left( 0\right) =0,$\ cannot be obtained by a recurrence. However,
the event $H\geq h\geq 1$\ is realized whenever a first birth event occurs
with size $\beta _{1}\geq h$\ or, if $\beta _{1}<h$, whenever for all states 
$h^{\prime }\geq h$ being hit when the amplitude $\beta $\ of a last upper
jump is larger than $h^{\prime }-h$, then $\tau _{\beta _{1},h^{\prime
}}<\tau _{\beta _{1},0}.$\ Hence, \newline

\textbf{Proposition.} \emph{For a recurrent walker with general }$\beta $%
\emph{, }$\mathbf{P}\left( H=\infty \right) =0$\emph{\ where, when }$h\geq
1, $%
\begin{eqnarray*}
\mathbf{P}\left( H\geq h\right) &=&p\mathbf{P}\left( \beta _{1}\geq h\right)
+p\sum_{x=1}^{h-1}\mathbf{P}\left( \beta _{1}=x\right) \sum_{h^{\prime }\geq
h}\frac{\varphi \left( x\right) }{\varphi \left( h^{\prime }\right) }\mathbf{%
P}\left( \beta >h^{\prime }-h\right) \\
&=&p\mathbf{P}\left( \beta _{1}\geq h\right) +p\cdot \sum_{x=1}^{h-1}\mathbf{%
P}\left( \beta _{1}=x\right) \varphi \left( x\right) \cdot \sum_{h^{^{\prime
\prime }}\geq 0}\frac{1}{\varphi \left( h+h^{\prime \prime }\right) }\mathbf{%
P}\left( \beta >h^{\prime \prime }\right)
\end{eqnarray*}
\emph{generalizing (\ref{h2}).}\newline

\section{Estimation from an $N$-sample of $X$, say $\left(
x_{0},x_{1},...,x_{N}\right) $}

We briefly sketch here how (in presence of real data which are suspected to
be in the binomial catastrophe framework), to estimate its constitutive
parameters.

From the transition matrix $P\left( x,y\right) $, the log-likelihood
function of the $N$-sample is 
\begin{equation*}
L\left( x_{0},x_{1},...,x_{N}\right) =\sum_{n=1}^{N}\left[ \log \left(
qd_{x_{n-1},x_{n}}\right) \mathbf{1}\left( x_{n}\leq x_{n-1}\right) +\log
\left( pb_{x_{n}-x_{n-1}}\right) \log \mathbf{1}\left( x_{n}>x_{n-1}\right)
\right] .
\end{equation*}
If one knows that some population grows and decays according to the binomial
catastrophe model with $\mathbf{E}\beta =\rho <\infty $, we propose the
following estimators: the Maximum-Likelihood-Estimator of $p$ while setting $%
\partial _{p}L=0$ is 
\begin{equation}
\widehat{p}=\frac{1}{N}\sum_{n=1}^{N}\mathbf{1}\left( x_{n}>x_{n-1}\right) .
\label{phat}
\end{equation}
With $\rho =\mathbf{E}\beta <\infty ,$ if the law of $\beta $ is a known
one-parameter $\rho -$family of pmfs, 
\begin{equation}
\widehat{\rho }=\frac{1}{\sum_{n=1}^{N}\mathbf{1}\left( x_{n}>x_{n-1}\right) 
}\sum_{n=1}^{N}\left( x_{n}-x_{n-1}\right) \mathbf{1}\left(
x_{n}>x_{n-1}\right) .  \label{rohat}
\end{equation}
Note that if 
\begin{equation*}
\widehat{\rho p}=\frac{1}{N}\sum_{n=1}^{N}\left( x_{n}-x_{n-1}\right) 
\mathbf{1}\left( x_{n}>x_{n-1}\right) ,
\end{equation*}
then $\widehat{\rho p}=\widehat{\rho }\widehat{p}.$

Also, in view of $\mathbf{P}\left( X_{n+1}=x\mid X_{n}=x\right) =qc^{x}$, $%
x\geq 0,$%
\begin{equation}
\widehat{c}=\frac{1}{\sum_{n=1}^{N}\mathbf{1}\left( x_{n}=1\right) }%
\sum_{n=1}^{N}\mathbf{1}\left( x_{n}=1,x_{n-1}=1\right) .  \label{chat}
\end{equation}

\section{Analysis of the absorbed version of $X_{n}$ and quasi-stationarity}

We consider again a version of $X_{n}$ started at $x_{0}>0,$ but which is
now absorbed when first hitting $0$. We aim at having additional insight
into the quasi-stationary distribution, which is known to be a difficult
issue. The absorbed process was considered in \cite{FS}. We work under the
condition that the original (non-absorbed) process is not transient, because
if it were, so would its absorbed version with a positive probability to
drift to $\infty $. The only thing which changes in the transition matrix $P$
as from (\ref{P}) is its first row which becomes $P\left( 0,y\right) =\delta
_{0,y}$, $y\geq 0$, leading to $P^{*}$. Letting $\Phi _{n}\left( z\right) =%
\mathbf{E}\left( z^{X_{n}}\right) $ with $\Phi _{0}\left( z\right)
=z^{x_{0}} $, taking into account the behavior of $X_{n}$ at $0$, with $\Phi
_{0}\left( z\right) =z^{x_{0}}$,the recurrence (\ref{R}) now becomes 
\begin{equation*}
\Phi _{n+1}\left( z\right) \overset{}{=\Phi _{n}\left( 0\right) +}pB\left(
z\right) \left( \Phi _{n}\left( z\right) -\Phi _{n}\left( 0\right) \right)
+q\left[ \Phi _{n}\left( 1-c\left( 1-z\right) \right) -\Phi _{n}\left(
0\right) \right] ,
\end{equation*}
\begin{eqnarray*}
\Phi _{n+1}\left( z\right) &=&p\left( 1-B\left( z\right) \right) \Phi
_{n}\left( 0\right) \overset{}{+}pB\left( z\right) \Phi _{n}\left( z\right)
+q\Phi _{n}\left( 1-c\left( 1-z\right) \right) , \\
\Phi _{n+1}\left( z\right) -\Phi _{n}\left( z\right) &=&p\left( B\left(
z\right) -1\right) \left( \Phi _{n}\left( z\right) -\Phi _{n}\left( 0\right)
\right) +q\left( \Phi _{n}\left( 1-c\left( 1-z\right) \right) -\Phi
_{n}\left( z\right) \right) .
\end{eqnarray*}
One can check that for all $z\in \left[ 0,1\right) ,$%
\begin{equation*}
\Phi _{\infty }\left( z\right) =1,\text{ consistently with }X_{\infty }=0%
\text{ (eventual extinction)}.
\end{equation*}
Note also 
\begin{eqnarray*}
\Phi _{n+1}\left( 1\right) &=&\Phi _{n}\left( 1\right) =\Phi _{0}\left(
1\right) =1\text{ (no mass loss),} \\
\Phi _{n+1}\left( 0\right) &=&p\Phi _{n}\left( 0\right) +q\Phi _{n}\left(
1-c\right) \text{ else,} \\
\Phi _{n+1}\left( 0\right) -\Phi _{n}\left( 0\right) &=&q\left( \Phi
_{n}\left( 1-c\right) -\Phi _{n}\left( 0\right) \right) >0\text{ if }n\geq 1,
\end{eqnarray*}
showing that $\Phi _{n}\left( 0\right) =\mathbf{P}\left( X_{n}=0\mid
X_{0}=x_{0}\right) $ is increasing tending to $1$ under the recurrence
condition. Finally,

- $\Phi _{n}\left( 0\right) =1\Leftrightarrow X_{n}=0\Leftrightarrow \Phi
_{n}\left( z\right) =1\Rightarrow \Phi _{n}\left( 1-c\right) =1\Rightarrow
\Phi _{n+1}\left( 0\right) =1$ ($0$ is absorbing)

- $X_{n}=0\Leftrightarrow \tau _{x_{0},0}\leq n\Rightarrow \Phi _{n}\left(
0\right) =\mathbf{P}\left( \tau _{x_{0},0}\leq n\right) $.

- $\mathbf{P}\left( \tau _{x_{0},0}=n\right) =\mathbf{P}\left( \tau
_{x_{0},0}\leq n\right) -\mathbf{P}\left( \tau _{x_{0},0}\leq n-1\right)
=\Phi _{n}\left( 0\right) -\Phi _{n-1}\left( 0\right) >0$.\newline

Here $\tau _{x_{0},0}$ be the first (and last) hitting time of $0$ for $%
X_{n} $ given $X_{0}=x_{0}>0$. It has the same distribution as the one
obtained for the original non-absorbed chain. Considering the substochastic
transition matrix $\overline{P}$ of $X_{n}$ where the first row and column
have been removed, the dynamics of 
\begin{equation*}
\Psi _{n}\left( z\right) =\mathbf{E}\left( z^{X_{n}}\mathbf{1}_{\tau
_{x_{0},0}>n}\right)
\end{equation*}
is 
\begin{equation}
\Psi _{n+1}\left( z\right) =pB\left( z\right) \Psi _{n}\left( z\right)
+q\left[ \Psi _{n}\left( 1-c\left( 1-z\right) \right) -\Psi _{n}\left(
1-c\right) \right] \text{, }\Psi _{0}\left( z\right) =z^{x_{0}}\text{, }\Psi
_{0}\left( 0\right) =0.  \label{Rpsi}
\end{equation}
Note $\Psi _{n}\left( 0\right) =0$ entails $\Psi _{n+1}\left( 0\right) =0:$
as required, there is no probability mass at $0$, because $\Psi _{n}\left(
z\right) $ is the pgf of $X_{n}$, on the event $\tau _{x_{0},0}>n$.

Furthermore, $\Psi _{n+1}\left( 1\right) =p\Psi _{n}\left( 1\right) +q\left[
\Psi _{n}\left( 1\right) -\Psi _{n}\left( 1-c\right) \right] =\Psi
_{n}\left( 1\right) -q\Psi _{n}\left( 1-c\right) ,$ so that $\Psi
_{n+1}\left( 1\right) -\Psi _{n}\left( 1\right) <0$ translating a natural
loss of mass.

We have $\Psi _{n}\left( 1\right) =\mathbf{P}\left( \tau _{x_{0},0}>n\right) 
$ (note $\Psi _{n}\left( 1\right) =1-\Phi _{n}\left( 0\right) $).
Conditioning, with $\Phi _{n}^{c}\left( z\right) =\mathbf{E}\left(
z^{X_{n}}\mid \tau _{x_{0},0}>n\right) ,$ upon normalizing, we get\newline

\textbf{Proposition.} \emph{Under the positive recurrence condition for }$%
X_{n}$\emph{, as }$n\rightarrow \infty $%
\begin{equation}
\Phi _{n}^{c}\left( z\right) :=\frac{\Psi _{n}\left( z\right) }{\Psi
_{n}\left( 1\right) }\rightarrow \frac{\Psi _{\infty }\left( z\right) }{\Psi
_{\infty }\left( 1\right) },  \label{qsdlim}
\end{equation}
\emph{the pgf of the quasi-stationary distribution }$\mathbf{l}$\emph{\ (as
a left eigenvector of }$\overline{P}$\emph{). An explicit expression of }$%
\Psi _{n}\left( z\right) $ \emph{follows from (\ref{psi}), (\ref{psi'})
below.}

The double generating function of $\Psi _{n}\left( z\right) $ is 
\begin{equation*}
\frac{1}{u}\left( \Psi \left( u,z\right) -z^{x_{0}}\right) =pB\left(
z\right) \Psi \left( u,z\right) +q\Psi \left( u,1-c\left( 1-z\right) \right)
-q\Psi \left( u,1-c\right)
\end{equation*}
Its iterated version is 
\begin{equation}
\Psi \left( u,z\right) =\sum_{n\geq 0}\left( qu\right) ^{n}\left(
1-c^{n}\left( 1-z\right) \right) ^{x_{0}}\prod_{m=0}^{n}H\left(
u,1-c^{m}\left( 1-z\right) \right)  \label{psi}
\end{equation}
\begin{equation*}
-\Psi \left( u,1-c\right) \sum_{n\geq 1}\left( qu\right)
^{n}\prod_{m=0}^{n-1}H\left( u,1-c^{m}\left( 1-z\right) \right) .
\end{equation*}
If $z=1-c$, 
\begin{eqnarray*}
\Psi \left( u,1-c\right) &=&\sum_{n\geq 0}\left( qu\right) ^{n}\left(
1-c^{n+1}\right) ^{x_{0}}\prod_{m=0}^{n}H\left( u,1-c^{m+1}\right) \\
&&-\Psi \left( u,1-c\right) \sum_{n\geq 1}\left( qu\right)
^{n}\prod_{m=0}^{n-1}H\left( u,1-c^{m+1}\right) ,
\end{eqnarray*}
so that 
\begin{equation}
\Psi \left( u,1-c\right) =\frac{\sum_{n\geq 0}\left( qu\right) ^{n}\left(
1-c^{n+1}\right) ^{x_{0}}\prod_{m=0}^{n}H\left( u,1-c^{m+1}\right) }{%
1+\sum_{n\geq 1}\left( qu\right) ^{n}\prod_{m=1}^{n}H\left( u,1-c^{m}\right) 
}.  \label{psi'}
\end{equation}
Plugging (\ref{psi'}) into (\ref{psi}) yields a closed form expression of $%
\Psi \left( u,z\right) $ and then of 
\begin{equation*}
\frac{\Psi _{n}\left( z\right) }{\Psi _{n}\left( 1\right) }=\frac{\left[
u^{n}\right] \Psi \left( u,z\right) }{\left[ u^{n}\right] \Psi \left(
u,1\right) }.
\end{equation*}
\newline

The value of $\Psi \left( u,z\right) $ at $z=1$ is $\Psi \left( u,1\right)
=\sum_{n\geq 0}u^{n}\mathbf{P}\left( \tau _{x_{0},0}>n\right) $ so that $%
\Psi \left( u,1\right) =\left( 1-\mathbf{E}u^{\tau _{x_{0},0}}\right)
/\left( 1-u\right) $. With $H\left( u,1\right) =1/\left( 1-pu\right) ,$ from
(\ref{psi}) 
\begin{eqnarray*}
\Psi \left( u,1\right) &=&\sum_{n\geq 0}\left( qu\right)
^{n}\prod_{m=0}^{n}H\left( u,1\right) -\Psi \left( u,1-c\right) \sum_{n\geq
1}\left( qu\right) ^{n}\prod_{m=0}^{n-1}H\left( u,1\right) \\
&=&\frac{1}{1-pu}\frac{1}{1-qu/\left( 1-pu\right) }-\Psi \left( u,1-c\right) 
\frac{qu}{1-pu}\frac{1}{1-qu/\left( 1-pu\right) } \\
&=&\frac{1}{1-u}-\Psi \left( u,1-c\right) \frac{qu}{1-u}=\frac{1}{1-u}\left(
1-qu\Psi \left( u,1-c\right) \right) .
\end{eqnarray*}
The pgf of $\tau _{x_{0},0}$ is thus also (a third representation of $\phi
_{x_{0},0}\left( u\right) :=\mathbf{E}u^{\tau _{x_{0},0}},$ see (\ref{R1})
and (\ref{R2})) 
\begin{equation}
\phi _{x_{0},0}\left( u\right) =qu\Psi \left( u,1-c\right) ,  \label{rep3}
\end{equation}
with $\Psi \left( u,1-c\right) $ given by (\ref{psi'}). This expression of $%
\phi _{x_{0},0}\left( u\right) $ is consistent with (\ref{taux00}).

\section{A variant of the binomial catastrophe model}

Suppose we are interested in the following simple semi-stochastic
decay/surge model: at each step of its evolution, the size of some
population either grows by a random number of individuals or shrinks by only 
\textbf{one} unit. The above binomial catastrophe model is not able to
represent this scenario where at least one individual is removed from the
population at catastrophic events. To remedy this, we therefore define and
study a variant of the above binomial model whereby the transition
probabilities in the bulk and at $0$ are slightly modified in order to
account for the latter decay/surge situation, \cite{EK2009}.

- If $X_{n}\geq 1$, define 
\begin{eqnarray*}
X_{n+1} &=&1\circ X_{n}+\beta _{n+1}=X_{n}+\beta _{n+1}\text{ wp }p \\
X_{n+1} &=&c\circ \left( X_{n}-1\right) \text{ wp }q
\end{eqnarray*}
Given a move down wp $q$: \textbf{One} individual out of $X_{n}$ is \textbf{%
systematically} removed from the population ($X_{n}\rightarrow X_{n}-1$);
each individual among the $X_{n}-1$ remaining ones being independently
subject to a survival/death issue (wp $c$ $/$ $1-c$) in the next generation.

- If $X_{n}=0$ ($p_{0}+q_{0}=1$) then, 
\begin{eqnarray*}
X_{n+1} &=&\beta _{n+1}\text{ wp }p_{0}, \\
&=&0\text{ wp }q_{0}.
\end{eqnarray*}
Unless $p_{0}=p$, our model yields some additional control on the future of
the population once it hits $0$ (extinction event).

With $b_{x}:=\mathbf{P}\left( \beta =x\right) $, $x\geq 1,$ the
one-step-transition matrix $P$ of the modified MC $X_{n}$ now is given by:

\begin{eqnarray*}
P\left( 0,0\right) &=&q_{0}\text{, }P\left( 0,y\right) =p_{0}\delta _{1}%
\text{, if }y\geq 1, \\
P\left( x,x\right) &=&0\text{ if }x\geq 1, \\
P\left( x,y\right) &=&q\binom{x-1}{y}c^{y}\left( 1-c\right) ^{x-1-y}\text{,
if }x\geq 1\text{ and }0\leq y<x, \\
P\left( x,y\right) &=&p\mathbf{P}\left( \beta =y-x\right) =pb_{y-x}\text{,
if }x\geq 1\text{ and }y>x.
\end{eqnarray*}
Note that because at least one individual dies out in a shrinking event, the
diagonal terms $P\left( x,x\right) $ of $P$ are now $0$ for all $x\geq 1.$%
\newline

\textbf{Remark}. \textit{One can introduce a holding probability }$r_{x}$%
\textit{\ to stay in state }$x$\textit{\ given }$X_{n}=x$, \textit{filling
up now the diagonal of} $P$\textit{. This corresponds to a time change while
considering a modified transition matrix }$\widetilde{P}$\textit{\ where: }$%
p\rightarrow p_{x}=p\left( 1-r_{x}\right) $\textit{\ and }$q\rightarrow
q_{x}=q\left( 1-r_{x}\right) $\textit{\ (}$p_{x}+q_{x}+r_{x}=1$\textit{).
So, with }$\mathbf{\rho :}=\mathbf{1}-\mathbf{r}$\textit{, a column vector
with entries }$\rho _{x}=1-r_{x}$, 
\begin{equation*}
P\rightarrow \widetilde{P}=I+D_{\mathbf{\rho }}\left( P-I\right) .
\end{equation*}
\textit{If the invariant measure }$\mathbf{\pi }$\textit{\ obeying }$\mathbf{%
\pi }^{\prime }=\mathbf{\pi }^{\prime }P,$\textit{\ (the fixed point of }$%
\mathbf{\pi }_{n+1}^{\prime }=\mathbf{\pi }_{n}^{\prime }P$\textit{) exists,
then the one of }$\widetilde{P}$\textit{\ also exists and obeys: }$%
\widetilde{\mathbf{\pi }}^{\prime }D_{\mathbf{\rho }}=\mathbf{\pi }^{\prime
} $\textit{.}

Suppose first $X_{0}=0$. Let $\mathbf{E}z^{X_{n}}:=\Phi _{n}\left( z\right) =%
\overline{\Phi }_{n}\left( z\right) +\Phi _{n}\left( 0\right) $ with $\Phi
_{0}\left( z\right) =1$ translating $X_{0}=0.$ Then, with $p+q=1$,

\begin{eqnarray*}
\Phi _{n+1}\left( z\right) &=&\left( q_{0}+p_{0}z\right) \Phi _{n}\left(
0\right) +pB\left( z\right) \overline{\Phi }_{n}\left( z\right) +\frac{q}{%
1-c\left( 1-z\right) }\overline{\Phi }_{n}\left( 1-c\left( 1-z\right)
\right) \text{, }\overline{\Phi }_{0}\left( z\right) =0, \\
\Phi _{n+1}\left( 0\right) &=&q_{0}\Phi _{n}\left( 0\right) +\frac{q}{1-c}%
\overline{\Phi }_{n}\left( 1-c\right) \text{, }\Phi _{0}\left( 0\right) =1.
\end{eqnarray*}
Note $\Phi _{n}\left( 0\right) =\mathbf{P}\left( X_{n}=0\right) $ and $%
\overline{\Phi }_{n}\left( 0\right) =0$ for each $n\geq 0.$

Thus, if these fixed point quantities exist

\begin{eqnarray*}
\overline{\Phi }_{\infty }\left( z\right) &=&p_{0}\left( z-1\right) \Phi
_{\infty }\left( 0\right) +pB\left( z\right) \overline{\Phi }_{\infty
}\left( z\right) +\frac{q}{1-c\left( 1-z\right) }\overline{\Phi }_{\infty
}\left( 1-c\left( 1-z\right) \right) , \\
\Phi _{\infty }\left( 0\right) &=&\frac{q}{p_{0}\left( 1-c\right) }\overline{%
\Phi }_{\infty }\left( 1-c\right) .
\end{eqnarray*}
We shall iterate the first fixed point equation which makes sense only when $%
c\neq 0$.

With $C\left( z\right) =\frac{p_{0}\left( z-1\right) }{1-pB\left( z\right) }$
and $D\left( z\right) =\frac{q}{1-c\left( 1-z\right) }\frac{1}{1-pB\left(
z\right) }$, we get 
\begin{eqnarray*}
\overline{\Phi }_{\infty }\left( z\right) &=&\Phi _{\infty }\left( 0\right)
\sum_{n\geq 0}C\left( 1-c^{n}\left( 1-z\right) \right)
\prod_{m=0}^{n-1}D\left( 1-c^{m}\left( 1-z\right) \right) \\
&=&\Phi _{\infty }\left( 0\right) \left[ C\left( z\right) +D\left( z\right)
\sum_{n\geq 1}C\left( 1-c^{n}\left( 1-z\right) \right)
\prod_{m=1}^{n-1}D\left( 1-c^{m}\left( 1-z\right) \right) \right] \\
&=&\frac{p_{0}\Phi _{\infty }\left( 0\right) \left( z-1\right) }{1-pB\left(
z\right) }\left[ 1+\sum_{n\geq 1}\left( cq\right) ^{n}\prod_{m=1}^{n}\frac{1%
}{1-c^{m}\left( 1-z\right) }\frac{1}{1-pB\left( 1-c^{m}\left( 1-z\right)
\right) }\right] .
\end{eqnarray*}
Except when $c=1$, the term inside the bracket has no pole at $z=1$. Then $%
\overline{\Phi }_{\infty }\left( 1\right) =0$ and so, assuming $\overline{%
\Phi }_{\infty }\left( 0\right) =0$, $\overline{\Phi }_{\infty }\left(
z\right) =1$ for all $z\in \left[ 0,1\right] $ is the only possible solution
to the first fixed point equation. Recalling from the second one that $\Phi
_{\infty }\left( 0\right) =\frac{q}{p_{0}\left( 1-c\right) }\overline{\Phi }%
_{\infty }\left( 1-c\right) $, we conclude that $\Phi _{\infty }\left(
0\right) =0$ and so $\Phi _{\infty }\left( z\right) =0$ for all $z\in \left[
0,1\right] .$ The only solution $\Phi _{\infty }\left( z\right) $ is the
trivial null one.\newline

It remains to study the cases $c=1$ and $c=0.$\newline

$\bullet $ If $c=1$. In this case, only a single individual can stepwise be
removed from the population; the transition matrix $P$ is upper- Hessenberg.
This constitutes a simple discrete version of a decay/surge model (some kind
of time-reversed version of the simple growth/collapse model). 
\begin{eqnarray*}
\overline{\Phi }_{\infty }\left( z\right) &=&\frac{p_{0}z\left( z-1\right) }{%
z\left( 1-pB\left( z\right) \right) -q}\Phi _{\infty }\left( 0\right) \\
&=&\frac{p_{0}z\left( z-1\right) }{z-1+p\left( 1-zB\left( z\right) \right) }%
\Phi _{\infty }\left( 0\right) .
\end{eqnarray*}
with $\overline{\Phi }_{\infty }\left( 0\right) =0.$ Letting $\overline{B}%
\left( z\right) =\left( 1-B\left( z\right) \right) /\left( 1-z\right) $ the
tail generating function of $\beta $, this is also 
\begin{eqnarray*}
\overline{\Phi }_{\infty }\left( z\right) &=&\frac{p_{0}z}{1-p\left( 1+z%
\overline{B}\left( z\right) \right) }\Phi _{\infty }\left( 0\right) \\
\Phi _{\infty }\left( z\right) &=&\left( 1+\frac{p_{0}z}{1-p\left( 1+z%
\overline{B}\left( z\right) \right) }\right) \Phi _{\infty }\left( 0\right) .
\end{eqnarray*}
With $\overline{B}\left( 1\right) =B^{\prime }\left( 1\right) =\mathbf{E}%
\beta =:\rho $, $\overline{\Phi }_{\infty }\left( 1\right) =\frac{p_{0}}{%
q-p\rho }\Phi _{\infty }\left( 0\right) =\frac{1}{1-c}\overline{\Phi }%
_{\infty }\left( 1-c\right) .$\newline

We conclude :

\textbf{Proposition }(phase transition)\textbf{.}

\emph{- Subcritical case: }$\overline{\Phi }_{\infty }\left( 1\right) +\Phi
_{\infty }\left( 0\right) =1\Rightarrow \Phi _{\infty }\left( 0\right)
=\left( q-p\rho \right) /\left( q-p\rho +p_{0}\right) $\emph{, well-defined
as a probability only if }$p\rho <q.$\emph{\ In this case, the chain is
positive-recurrent. The term }$p\rho $\emph{\ is the average size of a
move-up which has to be smaller than the average size }$q$\emph{\ of a
move-down.}

\emph{- Critical case: If }$q=p\rho $\emph{, then }$\Phi _{\infty }\left(
0\right) =0\Rightarrow \overline{\Phi }_{\infty }\left( z\right) =0$\emph{\
for each }$z$\emph{. The chain is null-recurrent and it has no non-trivial (}%
$\neq \mathbf{0}$\emph{) invariant measure }$\mathbf{\pi }.$

\emph{- Supercritical case: If }$\infty \geq p\rho >q$\emph{, the chain is
transient at }$\infty $\emph{.}\newline

\textbf{Examples:}

$\left( i\right) $ In addition to $c=1$, assume $B\left( z\right) =\overline{%
\alpha }z/\left( 1-\alpha z\right) $ (a geometric model for $\beta $ with
success probability $\overline{\alpha }$). Then, if $p<q\overline{\alpha }$, 
$\Phi _{\infty }\left( 0\right) =\pi \left( 0\right) =\left( q\overline{%
\alpha }-p\right) /\left( \overline{\alpha }\left( q+p_{0}\right) -p\right) $
is a probability and 
\begin{equation*}
\overline{\Phi }_{\infty }\left( z\right) =\frac{p_{0}z}{1-p\left(
1+z/\left( 1-\alpha z\right) \right) }\Phi _{\infty }\left( 0\right) =\frac{%
p_{0}z\left( 1-\alpha z\right) }{q-z\left( \alpha +p\overline{\alpha }%
\right) }\Phi _{\infty }\left( 0\right) .
\end{equation*}
Thus, with ($\alpha <a:=\left( \alpha +p\overline{\alpha }\right) /q<1$), we
obtain that 
\begin{equation*}
\pi \left( x\right) =\left[ z^{x}\right] \overline{\Phi }_{\infty }\left(
z\right) =\pi \left( 0\right) \frac{p_{0}}{q}\left( a-\alpha \right) a^{x-2}%
\text{, }x\geq 1
\end{equation*}
is the invariant probability measure of the chain, displaying geometric
decay at rate $a$.

$\left( ii\right) $ If in addition to $c=1$, we assume $B\left( z\right) =z$%
, $X$ is reduced to a simple birth and death chain (random walk) on the
non-negative integers, reflected at the origin. In this case, we get $%
\overline{\Phi }_{\infty }\left( z\right) =\frac{p_{0}z}{1-p\left(
1+z\right) }\Phi _{\infty }\left( 0\right) $ with ($\rho =1$): $\Phi
_{\infty }\left( 0\right) =\left( q-p\right) /\left( q-p+p_{0}\right) .$ The
corresponding chain is positive recurrent if $p<1/2$, null-recurrent if $%
p=1/2$ and transient at $\infty $ when $p>1/2.$ In the positive-recurrent
case, with $\pi \left( 0\right) =\left( 1-2p\right) /\left(
1-2p+p_{0}\right) $, we have 
\begin{equation*}
\pi \left( x\right) =\left[ z^{x}\right] \overline{\Phi }_{\infty }\left(
z\right) =\pi \left( 0\right) \frac{p_{0}}{q}\left( \frac{p}{q}\right) ^{x-1}%
\text{, }x\geq 1,
\end{equation*}
a well-known result, \cite{Feller}.

In both examples, whenever the process is positive recurrent, the invariant
measure having geometric decay at different rates $a$ and $p/q<1$
respectively.\newline

$\bullet $ If $c=0$ (total disasters) 
\begin{eqnarray*}
\overline{\Phi }_{\infty }\left( z\right) &=&p_{0}\left( z-1\right) \Phi
_{\infty }\left( 0\right) +pB\left( z\right) \overline{\Phi }_{\infty
}\left( z\right) +q\overline{\Phi }_{\infty }\left( 1\right) \\
\Phi _{\infty }\left( 0\right) &=&\frac{q}{p_{0}}\overline{\Phi }_{\infty
}\left( 1\right) ,
\end{eqnarray*}
leading to $\Phi _{\infty }\left( 0\right) =q/\left( p_{0}+q\right) =\pi
\left( 0\right) $ and 
\begin{equation*}
\overline{\Phi }_{\infty }\left( z\right) =\frac{p_{0}z}{1-pB\left( z\right) 
}\Phi _{\infty }\left( 0\right) ,\text{ or}
\end{equation*}
\begin{equation}
\Phi _{\infty }\left( z\right) =\left( 1+\frac{p_{0}z}{1-pB\left( z\right) }%
\right) \Phi _{\infty }\left( 0\right) .  \label{pgftd}
\end{equation}
We conclude:

\textbf{Proposition.} \emph{In the total disaster case, the modified chain
is always positive recurrent with invariant measure having pgf (\ref{pgftd}%
), as a shifted compound geometric distribution.}

\textbf{Examples: }

$\left( i\right) $ In addition to $c=0$, assume $B\left( z\right) =\overline{%
\alpha }z/\left( 1-\alpha z\right) $ (a geometric model for $\beta $ with
success probability $\overline{\alpha }$). Then 
\begin{equation*}
\Phi _{\infty }\left( z\right) =\left( 1+\frac{p_{0}z}{1-pB\left( z\right) }%
\right) \Phi _{\infty }\left( 0\right) .
\end{equation*}

$\left( ii\right) $ In addition to $c=0$, assume $B\left( z\right) =z.$
Then, with $\pi \left( 0\right) =q/\left( p_{0}+q\right) ,$%
\begin{eqnarray*}
\Phi _{\infty }\left( z\right) &=&\left( 1+\frac{p_{0}z}{1-pz}\right) \Phi
_{\infty }\left( 0\right) , \\
\pi \left( x\right) &=&\left[ z^{x}\right] \Phi _{\infty }\left( z\right)
=\pi \left( 0\right) p_{0}p^{x-1}\text{, }x\geq 1,
\end{eqnarray*}
a geometric distribution with decay rate $p$.\newline

\textbf{Acknowledgments:} T. Huillet acknowledges partial support from the
``Chaire \textit{Mod\'{e}lisation math\'{e}matique et biodiversit\'{e}''.}
B. Goncalves and T. Huillet acknowledge support from the labex MME-DII
Center of Excellence (\textit{Mod\`{e}les math\'{e}matiques et
\'{e}conomiques de la dynamique, de l'incertitude et des interactions},
ANR-11-LABX-0023-01 project). Finally, this work was also funded by CY
Initiative of Excellence (grant ``Investissements d'Avenir'' ANR-
16-IDEX-0008), Project EcoDep PSI-AAP 2020-0000000013. We warmly thank our
Referees for suggesting significant improvements in the presentation of this
manuscript.

\end{document}